\documentclass[%
 reprint,
unsortedaddress,
 amsmath,amssymb,
 aps,
pra,
]{revtex4-1}

\usepackage{xcolor}
\usepackage{graphicx}% Include figure files
\usepackage{dcolumn}% Align table columns on decimal point
\usepackage{bm}% bold math
\usepackage{comment}
\usepackage{hyperref}% add hypertext capabilities
\usepackage{qcircuit}
\usepackage{tikz}
\usepackage{braket}
\begin{document}

\preprint{APS/123-QED}

\title{
Non-unitary operations for ground-state calculations in near term quantum computers
}% Force line breaks with \\

\author{Guglielmo Mazzola}
\affiliation{IBM Research Zurich, S\"aumerstrasse 4, 8803 R\"uschlikon, Switzerland}
\email{GMA@zurich.ibm.com}
\author{Pauline Ollitrault}
\affiliation{IBM Research Zurich, S\"aumerstrasse 4, 8803 R\"uschlikon, Switzerland}
\affiliation{Laboratory of Physical Chemistry,
ETH Z\"urich, 8093 Z\"urich, Switzerland}
\author{Panagiotis Kl. Barkoutsos}
\affiliation{IBM Research Zurich, S\"aumerstrasse 4, 8803 R\"uschlikon, Switzerland}
\author{Ivano Tavernelli}
\affiliation{IBM Research Zurich, S\"aumerstrasse 4, 8803 R\"uschlikon, Switzerland}

\date{\today}% It is always \today, today,
             %  but any date may be explicitly specified

\begin{abstract}

We introduce a quantum Monte Carlo inspired reweighting scheme  to accurately compute energies from optimally short quantum circuits.
 This effectively hybrid  quantum-classical approach  features both  entanglement provided by a short quantum circuit, and the presence of an effective non-unitary operator at the same time. 
 The functional form of this projector is borrowed from classical computation and is able to filter-out high-energy components generated by a sub-optimal variational quantum heuristic ansatz.
 The accuracy of this approach is demonstrated  numerically in finding energies of entangled ground-states of many-body lattice models.
 We demonstrate a practical implementation on IBM quantum hardwares up to an 8 qubits circuit.

\end{abstract}

\pacs{Valid PACS appear here}
\maketitle

\section{Introduction}

Solving quantum many-body and electronic structure problems is one of the most anticipated applications of quantum computers, in view of the exponential speed-up that can be achieved compared to classical simulations\cite{Feynman1982,lanyon2010towards,PhysRevA.92.062318}.
Despite decades of efforts, an efficient classical way to describe many-body effects and strong correlations is still missing, preventing classical computation of fermionic systems from reaching the desired accuracy in large-scale applications\cite{PhysRevLett.94.170201,Becca2017}.
On the other hand, quantum computation is still at its infancy and state-of-the-art calculations are performed on so-called noisy intermediate quantum (NISQ) hardware, of 20-50 qubits\cite{preskill18}.

These non-ideal conditions, represented by short circuit depths and the absence of implementable error correction schemes, call for the development of suitable algorithms able to exploit the present resources\cite{Moll2018,Cross2018}.
In this context, variational approaches have been proposed as near-term strategy to solve the electronic structure problem\cite{peruzzo2014variational,kivlichan2018quantum,shen_quantum_2017,PhysRevA.92.042303}.
These algorithms drastically reduce the coherence time requirement, but feature
optimizable parameters $\bm{\theta}$ in the circuit, generating a parametrized quantum state $ | \psi_c (\bm{\theta}) \rangle$.
These parameters are optimized to minimize the energy $\langle \psi_c (\bm{\theta})| \mathcal{H} | \psi_c (\bm{\theta})\rangle$ for a given problem Hamiltonian $\mathcal{H}$.
The energy is calculated as a sum of expectation values of Pauli operators, hence the circuit is executed multiple times to reduce the variance of such estimates. The parameter optimization is instead performed classically\cite{peruzzo2014variational}. This approach, called
 variational quantum eigensolver (VQE),  has been applied to small molecules and quantum magnets\cite{omalley_scalable_2016,kandala_hardware-efficient_2017,Ganzhorn2018,Hempel2018}, and  relies on the assumption that a quantum state prepared in a quantum computer can represent efficiently and compactly all the correlations that are hard to encode classically\cite{PhysRevLett.108.110502}.

\begin{figure*}[hbt]
\includegraphics[width=1.0\textwidth]{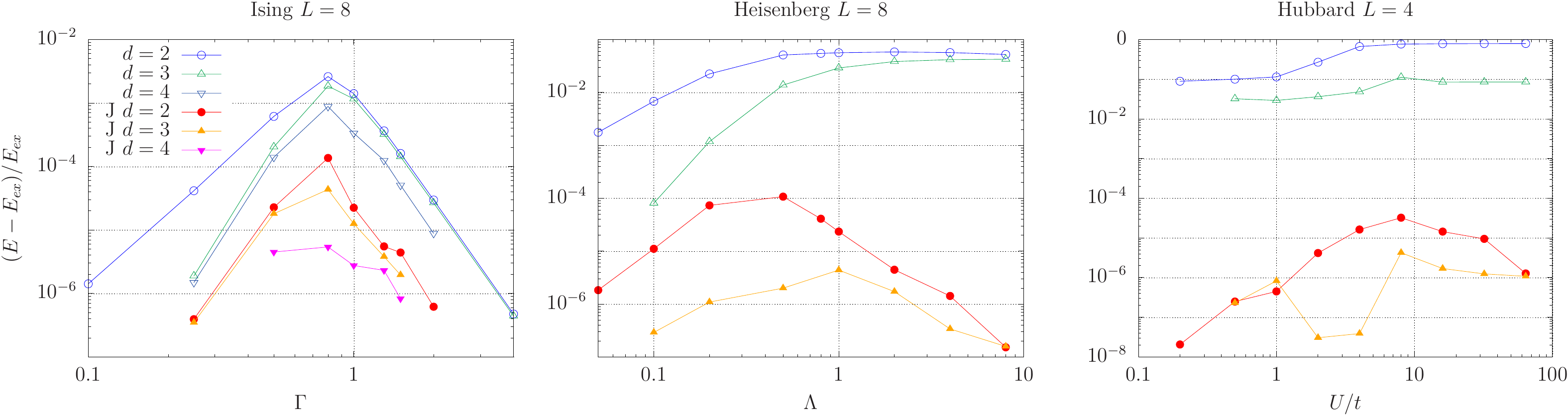}
\caption{
\emph{Accuracy of the JQC state for one-dimensional many-body problems}. Relative energy differences are shown as a function of the relevant Ising, Heisenberd and Hubbard model parameters.
Energies computed for selected circuit depths $d$ are plotted for the circuit ansatz (empty symbols, blue and green) and for the JQC (solid, red and orange).
All systems sizes translate into a $N=8$ qubit register.
The $R_y$CNOT circuit ansatz is lacking any a-priory design based on physical intuition. While the JQC ansatz always improves upon the circuit one, notably the worst performance is around the critical points of the models, i.e. when $\Gamma,\Lambda, U/4t=1$ respectively.
The number of Jastrow optimizable variational parameters, exploiting lattice symmetries, is $7$ in case of Ising and Heisenberg models, and $10$ for Hubbard, since here the fermion-to-qubit mapping implies a ladder structure (see Supplementary Materials\cite{supmat}). For each setup we plot the best outcome amongst several repetions of the numerical optimization procedure.
}
\label{fig:result}
\end{figure*}

Lattice many-body models represent an ideal testbed for developing new algorithms, since they retain all the features that make electronic structure problems hard to simulate classically, but without the specific overcomplication of quantum chemistry (i.e. the generation of the Hamiltonian parameters that always require a classical preprocessing tool).

A concrete example is the Hubbard model, which is perhaps the most extensively studied condensed matter system, as it serves as a minimal model for high-temperature superconductors\cite{RevModPhys.66.763} and other correlation-driven phase transitions\cite{Dagotto257}.
The exponential scaling of the Hilbert space's size with respect to system size $L$ prevents polynomially scaling classical algorithms from an accurate solution in most of the cases, except from particularly symmetric conditions such as two-dimensional (2D) lattices at half-filling\cite{Becca2017}.

The most advanced classical algorithms,  such as quantum Monte Carlo (QMC) or density matrix renormalization group (DMRG) theory\cite{PhysRevX.5.041041} are also characterized by underlying variational states.
Interestingly, it has been noted that
results may depend from the structure of the variational form used. An example is
 the debated existence of the so-called stripe order, which is a state displaying charge and spin modulations,  in the underdoped region of the 2D Hubbard model\cite{Zheng1155,PhysRevLett.113.046402,PhysRevLett.88.117002,PhysRevB.85.081110,note2}.

 Other examples concern the proposed spin-liquid character of the Heisenberg antiferromagnet on the Kagome lattice\cite{jiang2012identifying,PhysRevB.87.060405}, the variational description of frustrated spin models\cite{PhysRevLett.87.097201}, and Mott insulators\cite{PhysRevLett.94.026406}.

The need for an accurate and easy to prepare variational trial state is transferred in the realm of quantum computation.
In the VQE approach, the trial state's ability of describing the desired physical state is determined by the set of gates composing the quantum circuit and is limited by the affordable circuit depth. The connectivity between the qubits also plays an important role since the presence of at least a set of two-qubit gates is necessary to achieve a final entangled state.
Due to the limited coherence time of present NISQ machines, it is only possible to run relatively short circuits, with a detrimental impact on the accuracy of the calculation.
For example, the unitary coupled cluster (UCC) ansatz\cite{Watts1989}, which is the quantum counterpart of the celebrated coupled-cluster technique\cite{Bartlett1978}, has been proposed as a polynomially scaling quantum circuit to solve quantum chemistry problems.
However, the number of gates necessary to achieve chemical accuracy, even on small molecules, is simply too large to be successfully executed on NISQ devices\cite{wecker2014gate}.

Heuristic circuits, which implement hardware-efficient gates, represent a more realistic approach in the short term, and have been already demonstrated in several small chemical \cite{kandala_hardware-efficient_2017,Barkoutsos2018,Ganzhorn2018} and lattice model examples\cite{PhysRevA.94.032338}. However they suffer from the same coherence time limitation when investigating larger systems\cite{Barkoutsos2018}.

In this paper, we introduce a hybrid quantum-classical type of trial states $\mathcal{P}| \psi_c \rangle$, which exploits both the entanglement offered by a short quantum circuit, and projective pseudo-dynamics, implemented at the classical level, through measurements post-processing.
Here, the projector $\mathcal{P}$  filters out the unwanted high-energy components from the sub-optimal trial state  $| \psi_c \rangle$, produced by the circuit, and is inspired by established correlated  methods, such as  Variational Monte Carlo (VMC)\cite{Becca2017}.

\begin{figure*}[hbt]
\includegraphics[width=1.0\textwidth]{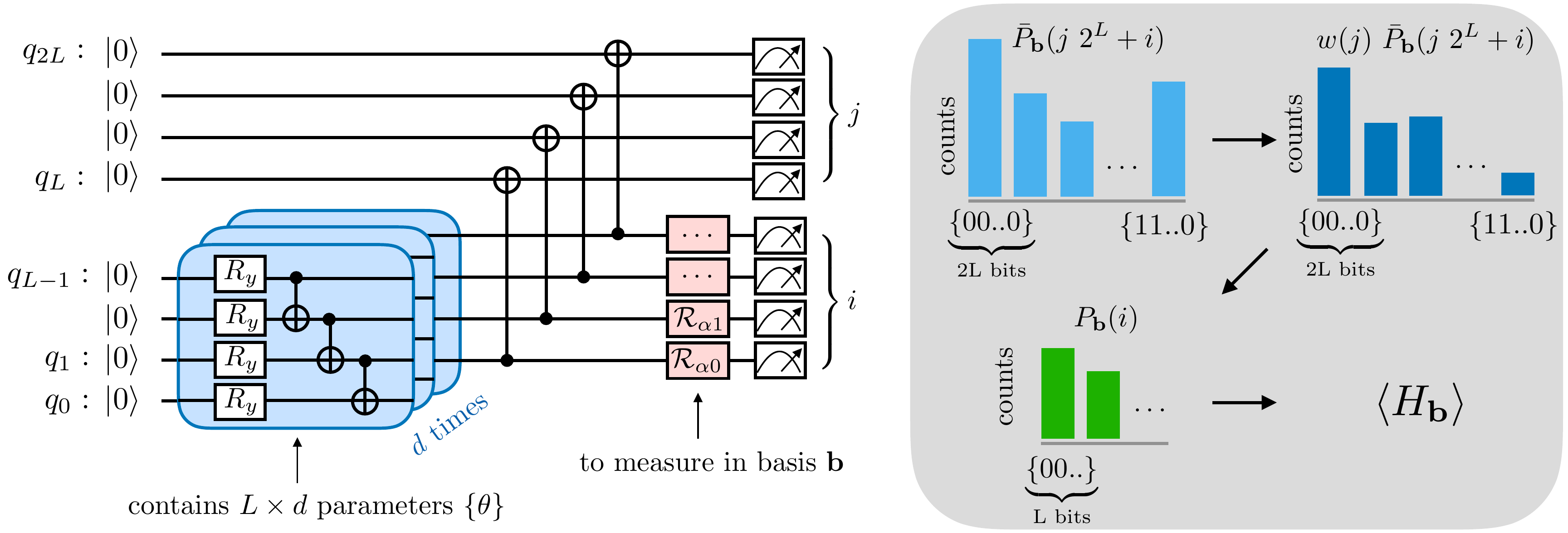}
\caption{
\emph{Hardware implementation and classical postprocessing.} \emph{Left.} Circuit realizing the entangled copy of the state produced by the variational block (blue). The most tested variational form in this work is the $R_y$ CNOT, made by $d$ repetitions of blocks. Each of them features $L$ parametrized single qubits rotations $R_y$, and a cascade of $L-1$ CNOT gates. Postrotations are applied only on $L$ qubits to measure outcomes in a given basis ${\bf b}$. \emph{Right.} For each basis ${\bf b}$, the normalized measured count $\bar P_{\bf_b}$ of the $2L$-bit possible outcomes (each of them univocally defined by the integer $2^L j+i$), is reweighted according to the Jastrow function. The probability distribution  $P_{\bf_b}$, defined on $L-$bit strings is then recovered, and used to evaluate the expectation value of $H_b$ (see main text).
}
\label{fig:circuit}
\end{figure*}

We propose two quite different practical approaches to implement the non-unitary operator  $\mathcal{P}$. 
%In the first, we use an additional qubit register.
In the first, the information stored in an ancillary register is used to re-weight the measurements performed on the $N$-qubit circuit register.
The second strategy does not require ancillary qubits but the evaluation of the modified Hamiltonian $\mathcal{P} \mathcal{H} \mathcal{P}$. Depending on the complexity of  $\mathcal{P}$, this translates in a polynomial increase of the number of Pauli terms to measure.

\section{Projectors in Quantum Monte Carlo}

 The  projector $\mathcal{P}$ (partially)  removes the residual components of the circuit ansatz $|\psi_c \rangle$  having negligible overlap with the target state (cfn. also the Supplementary Materials for an illustrative example\cite{supmat}).
The knowledge of the exact state is however not required to construct such operator.
To this end, we borrow inspiration from classical simulations, where
physically motivated classes of projectors have already demonstrated good accuracy in describing strong correlations.
These are the so-called Jastrow functions, widely used in the QMC community in solving lattice models\cite{RevModPhys.63.1,PhysRevB.77.024510}and continuous systems\cite{Jastrow}, in both first\cite{casula2004correlated,zen2014static, PhysRevLett.120.025701,genovese} and second quantization\cite{neu1,neu2}.

For example, a particularly simple but effective projector, the so-called Gutzwiller operator\cite{PhysRevLett.10.159}, counts the number of doubly occupied sites in a lattice, removing such high-energy components in the Hubbard model at large-$U$.
The same operator may as well suppress ionic terms naturally arising from a simple single-particle product state description of molecular dissociation.

\section{Jastrow quantum circuit states.}

The strategy we propose in this paper is to act directly on the $N$-qubits space e.g., in the case of the Hubbard model, after the Jordan-Wigner mapping of fermionic operator to the qubit space\cite{PhysRevA.92.062318,PhysRevA.94.032338}. 
A long-ranged spin Jastrow operator is then applied to the circuit ansatz, using the projector
\begin{equation}
   \mathcal{P}_J=e^{J},\quad J = \sum_{k,l=1 (k\ne l)}^N \lambda_{kl} \sigma^z_k \sigma^z_l\quad,
\label{eq:jastrow}
\end{equation}
where $\sigma^\alpha_k$ are Pauli matrices, and $\lambda_{k,l}$ are $N(N-1)/2$ variational parameters.
The number of effective optimizable paramenters $\bm \lambda$ can be reduced by applying lattice symmetries, for example by assuming that the value of $\lambda_{k,l}$ only depends on the distance between qubits $k$ and $l$.
The Jastrow correlator can be generalized also to three and more spin interactions.
For the sake of brevity, we propose the name \emph{Jastrow quantum circuit} (JQC) for the $\mathcal{P}_J | \psi_c \rangle$ state, reminiscent of the Jastrow Slater Determinant (JSD) wavefunctions used in electronic  QMC calculations\cite{Jastrow,RevModPhys.73.33}.
In our case the qubit Jastrow operator improves the description of spin correlations by acting on the circuit ansatz, whereas in the classical counterpart it is applied to a mean-field  starting state, which imposes the correct (anti)symmetrization of the system.
Moreover, it easily includes all possible two qubits $k,l$ interactions being not constrained by the available hardware connectivity\cite{PhysRevLett.120.110501}.

\subsection{Accuracy of the JQC variational states.}

We tested the accuracy of the JQC ansatz on three popular many-body models in one dimension: the transverse field Ising model,
$\mathcal{H}_{Ising} = -\sum_{k,l=0 (k\ne l)}^{L-1} \sigma^z_k \sigma^z_l + \Gamma \sum_{k=0}^{L-1} \sigma^x_k$, the Heisenberg model $\mathcal{H}_{Heis} = -\sum_{k,l=0 (k\ne l)}^{L-1} \sigma^z_k \sigma^z_l + \Lambda   ( \sigma^x_k \sigma^x_l +  \sigma^y_k \sigma^y_l )$, and the Hubbard model $\mathcal{H}_{Hub} = - t \sum_{k=0}^{L-1} \sum_{s=\uparrow,\downarrow} (c_{k,s}^{\dag} c_{k+1,s} + c_{k+1,s}^{\dag} c_{k,s}) +
    U \sum_{k=0}^{L-1} (c_{k,\uparrow}^{\dag} c_{k,\uparrow}  c_{k,\downarrow}^{\dag} c_{k,\downarrow})$
at half-filling, where $c_{k,s}^{\dag}$ ($c_{k,s}$) are fermionic creation(destruction) operators at site $i$, and $t$, $U$ are the hopping and on-site Coulomb repulsion parameters, respectively.
While the first two models do not require any mapping, being already spin Hamiltonians (therefore $N=L$), we use the mapping between spinful electrons and qubits, illustrated in Ref.\cite{verstraete2005mapping,PhysRevB.77.024510}, to map a $L$-sites Hubbard model into a $N=2L$ qubits register with ladder connectivity (cfn. also Supplementary Materials\cite{supmat}).

In this work we use a primitive heuristic $R_y$-CNOT circuit. 
%(cfn Fig.~\ref{fig:rycnot})..
The circuit ansatz is represented by  $| \psi_c({\bm \theta})  \rangle = U_{c}({\bm \theta}) | \psi_{init}\rangle$,
where $U_{c}({\bm \theta})$ is the unitary operator representing the circuit, ${\bm \theta}$ is the set of total $d~N$ single-qubit rotation angles, where $d$  is the circuit depth, and $|\psi_{init}\rangle$ is an easy-to-prepare bit-string.
While such type of circuit requires an affordable number of entangling gates (CNOT) per block,
it may not respect basic symmetries of the desired solution, compatible with the particle and spin number conservation\cite{Barkoutsos2018}.

\begin{figure}[hbt]
\includegraphics[width=1.0\columnwidth]{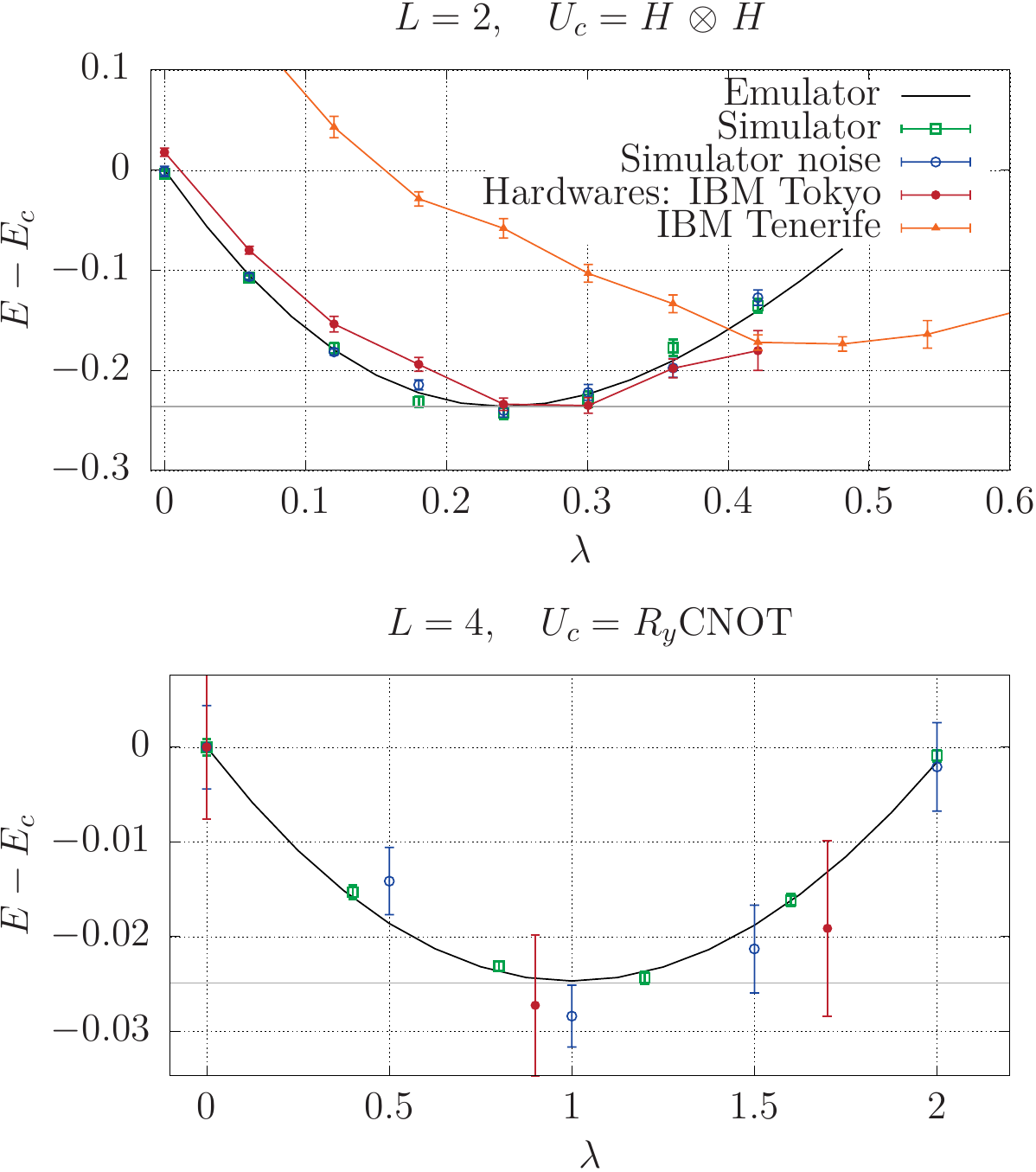}\caption{
\emph{JQC ansatz on hardware and circuit simulators.} \emph{Top panel.} Energy difference of the JQC states compared to the circuit energy $E_c$ for a $L=2$ Ising model as a function of the single Jastrow parameter $\lambda$ (cfn. main text).  The circuit considered is  $U_c = H \otimes H$. The black(gray) line represents the exact energy of the JQC ansatz(of the Ising model). Colored points are obtained implementing the extended circuit of Fig.~\ref{fig:circuit} with simulators (green = noiseless simulator, blue = noise model from IBM Tokyo chip) and with hardwares (red = IBM Tokyo chip, orange = IBM Tenerife), and using the measurements reweighting method introduced in the main text. For each $\lambda$ point we reconstruct the probabilities in the $ZZ$ and $XX$ basis using 8192 shots.
Error bars are computed repeating this process $M_{rep}=12$ times.
The JQC state reduce to $\psi_c$ for $\lambda=0$.
\emph{Bottom panel}. Same analysis as above, but on a $L=4$ system and using an entangled $R_y$CNOT circuit. Here the Jastrow operator contains 3 variational parameters, this set $\lambda_{opt}$ is optimized beforehand. In order to have an one-dimensional plot we multiply them elementwise by $\lambda$, such that when $\lambda=1$ we re-obtain the optimal solution.
In the noiseless simulator case we acquire $2\times10^6$ shots, with $M_{rep}=24$. In the noisy simulator(hardware) case we acquire $1.6\times10^5$($2.5\times10^4$) shots, and $M_{rep}=12(24)$. In the latter cases, a rigid shift of $\approx 0.15$ is applied to the data-series so that the $\lambda=0$ point is at $E-E_c=0$.
}
\label{fig:reshard}
\end{figure}

In this case, the non-unitary Jastrow operator will effectively project-out wave-function components of $|\psi_c \rangle$ having particle number and  magnetization incompatible with the  Heisenberg and Hubbard models.
More generally, $\mathcal{P}_J$ will provide a modulation to the amplitudes of the quantum state $|\psi_c \rangle$,
while preserving their sign.
For this reason, as it happens in VMC, the Jastrow operator cannot recover the exact energy by itself, irrespectively of the circuit.
Nevertheless, the JQC ansatz uniformly improves the standard VQE ansatz $| \psi_c({\bm \theta})  \rangle$, for all the three models considered.
In fact, as we observe a several order-of-magnitude improvement of the energy at fixed circuit depths (cfn. Fig~\ref{fig:result}).

As discussed above, the total number of variational parameters in the set $( {\bm \theta},{\bm \lambda} )$ is linearly increasing with $N$, if appropriate lattice symmetries are taken into account (cfn. Supplementary Materials\cite{supmat}), or if a cut-off is imposed on the correlation lenghts considered in Eq.~\ref{eq:jastrow}.

While in Fig.~\ref{fig:result} we benchmark the quality of the ansatz at fixed system sizes by varying the models parameters, in the Supplemetary Materials we investigate the efficiency of our method as a function of the system size\cite{supmat}.
These results have been obtained simulating the circuit in noiseless conditions and applying exactly the Jastrow operator on the state-vector $| \psi_c \rangle$.
Since the JQC state is not normalized to unity, the normalization has been computed numerically.

\section{Hardware implementation}

\subsection{ Entangled copy method } 

Differently to classical simulations, implementing the projector is not as straightforward. 
In this section we will illustrate the general procedure, while developing in parallel an example (a $L=2$ sites Ising model.)
The standard VQE approach  would require a 2 qubits circuit, and measuring 
 expectation values of the $ZZ$, $XI$, and $IX$ operators,  to calculate the energy of the model\cite{kandala_hardware-efficient_2017}.
 In practical approaches (as implemented in the Qiskit package\cite{Qiskit}) this is done by separating the Hamiltonian in groups of operator that can be measured simultaneously
 $H=\sum_{\bf b} H_{\bf b}$, in a given basis ${\bf b}$.
 The expectation value $\langle H_{\bf b} \rangle$ is evaluated by reconstructing the bit-string probability $P_{\bf_b}(i)$ in the ${\bf b}$ basis, through measurements (here the label $i$ denotes the integer encoded in the $L$-bit string)\cite{note3}.
  Suitable unitaries (called post-rotations $\mathcal{R}_{\alpha}$) allows us to measure in different bases. In this example, to  measure the non-diagonal operator we need $\mathcal{R}_{\alpha} = H$, where $H$ is the Hadamard matrix.

In our approach, we need to combine the possibility of measuring the qubits after applying the postrotations, with the requirements to also read the qubits in their $Z$ basis, to evaluate Eq.~\ref{eq:jastrow}.
This is possible only by introducing an ancillary register of the same size $L$.
We use the ancilla register to store an entangled copy of the original register, using CNOT gates as in Fig.~\ref{fig:circuit}.
In this specific case, the total register now reads $\{q_0,q_1.q_2,q_3 \}$, where the first $L=2$ qubits evolve through the circuit, and the last  ones are initialized to 0.
In such a way, the $L$-bit outcome $j$ of the measurement on the ancilla register, determines a weight $w(j)$, to be applied to the $2L$-bit readout, labelled with the integer $2^L j+i$. 
The only additional step required is to reconstruct the reduced probability of the original 2-bits components $P_{\bf_b}(i)$, given the total probability of the 4-bits strings $\bar P_{\bf_b}(2^L j+i)$, which is actually measured and then reweighted
(see Supplementary Material\cite{supmat}).

In Fig.~\ref{fig:reshard} we benchmark the proposed reweighting scheme on syntetic measurements (i.e. using circuit simulators, with and without noise), and real datasets from IBM Q hardwares, Tokyo (20 qubits) and Tenerife (5 qubits) using  $L=2$ and $L=4$ Ising models, and with two different circuit ansatzes.

The first circuit is made of two Hadamard gates, $U_c = H \otimes H$, that produces an equal superposition state 
%$\psi_c= 1/2 ( |00\rangle +  |01\rangle +  |10\rangle + |11\rangle )$.
Here, it can be shown analitically that a two-spins Jastrow operator (cfn. Eq.~\ref{eq:jastrow}) suffices to amplify(suppress) the components having even(odd) parity, recovering the (unnormalized) exact state, for $\lambda=\lambda_{01}\approx0.24$.
Notice also that in this limiting case the circuit state is a trivial product state, and the Jastrow operator recovers all the missing correlations characterizing the exact state.

The obtained energies from the circuit simulations and the Tokyo machine data are compatible with the predicted values at variuos $\lambda$, obtained by state-vector emulations of the process. 
We observe that for the Tenerife hardware data the one-parameter Jastrow operator is not sufficient to recover the exact energy of the model, because of the noise level of the hardware.
Nevertheless, the Jastrow operator allows us to improve the circuit energy, although for a different value of $\lambda$ compared to the theoretically predicted one. 

The technique is also demonstrated on a more challenging $L=4$ system (which translates into a $2L=8$ qubit register). Here  the Jastrow projector recovers the energy difference between an heuristic $R_y$-CNOT (with $d=1$ and optimized $\bm{\theta}$ parameters)  energy $E_c$ and the exact one.

\subsection{Transformed Hamiltonian method } 

The second implementation we propose  requires additional Pauli operators to be measured, instead of ancillas qubits. Computing the expectation value of the energy on the JQC state $E =  \langle \mathcal{P}_J \psi_c | \mathcal{H} | \mathcal{P}_J \psi_c \rangle /  \langle \mathcal{P}_J \psi_c  | \mathcal{P}_J \psi_c \rangle  $, it is equivalent to measure the ratio of $\mathcal{P}_J \mathcal{H} \mathcal{P}_J$ and $\mathcal{P}_J \mathcal{P}_J$ operators on $\psi_c$.
Unfortunately we notice that Eq.~\ref{eq:jastrow} results in an exponentially increasing number of Pauli operators with the register size.
However, a suitably truncated expansion $\mathcal{P}_J^{'}= 1 + J + J^2/2 + \cdots$, controlled by the smallness of the
parameters $\bm{\lambda}$, can still be effective while reducing the number of operators to polynomial scaling
(we report numerical benchmarks in the Supplementary Materials\cite{supmat}).

\section{Conclusions.~}

We introduce hybrid quantum-classical states to solve  many-body lattice models, drastically reducing the depth requirements of the heuristic circuits to reach target accuracy, by leveraging on the use of non-unitary operators\cite{motta2019quantum}.
 The approach is variational, and after full optimization of the Jastrow parameters the energy is always better or -in the worst case- equal to the one provided by the circuit ansatz.
Two practical schemes have been proposed to realise such states in present hardwares. The most promising one requires an additional ancillary register. This methods relies on measurements reweighting and allows for an exact implementation of the Jastrow correlation operator, while it does not increase the number of Pauli operators to be measured, that  already constitutes a drawback of standard VQE\cite{PhysRevA.92.042303}.
The approach has been demonstrated using an 8 qubits circuit in the IBM Tokyo chip, providing quantitative energetics for the testcase Ising model.
 We notice that this implementation extends the scope of the recently proposed stabilizer-VQE method\cite{mcardle2018error}, where measurements are simply discarded using  an error detection scheme. In this case, the projector allows us to improve the accuracy of the VQE ansatz, while also mitigating the possible errors.
A possible issue of these approaches may arise when the VQE ansatz and the exact state have negligible overlap. 
In our case, an increased statistical fluctuation of the energy  would be a fingerprint that most of the measurements are reweighted to zero.
While this issue is not present in the studied cases (cfn. Supplementary Materials\cite{supmat}), it will call for the development of synergic circuit and projector operators.
We anticipate the use of suitably modified projected circuit states  for solving  quantum chemistry problems.

\begin{acknowledgments}

{\bf Acknowledments.} We acknowledge discussions with L.Guidoni, F. Benfenati, N. Moll, and S. Sorella. The authors acknowledge financial support from the Swiss National Science Foundation (SNF) through the grant No. 200021-179312.
\end{acknowledgments}

\bibliographystyle{elsarticle-num}
\bibliography{ms}

\begin{thebibliography}{10}
\expandafter\ifx\csname url\endcsname\relax
  \def\url#1{\texttt{#1}}\fi
\expandafter\ifx\csname urlprefix\endcsname\relax\def\urlprefix{URL }\fi
\expandafter\ifx\csname href\endcsname\relax
  \def\href#1#2{#2} \def\path#1{#1}\fi

\bibitem{Feynman1982}
R.~P. Feynman, \href{https://doi.org/10.1007/BF02650179}{Simulating physics
  with computers}, International Journal of Theoretical Physics 21~(6) (1982)
  467--488.
\newblock \href {http://dx.doi.org/10.1007/BF02650179}
  {\path{doi:10.1007/BF02650179}}.
\newline\urlprefix\url{https://doi.org/10.1007/BF02650179}

\bibitem{lanyon2010towards}
B.~P. Lanyon, J.~D. Whitfield, G.~G. Gillett, M.~E. Goggin, M.~P. Almeida,
  I.~Kassal, J.~D. Biamonte, M.~Mohseni, B.~J. Powell, M.~Barbieri, et~al.,
  Towards quantum chemistry on a quantum computer, Nature chemistry 2~(2)
  (2010) 106.

\bibitem{PhysRevA.92.062318}
D.~Wecker, M.~B. Hastings, N.~Wiebe, B.~K. Clark, C.~Nayak, M.~Troyer,
  \href{https://link.aps.org/doi/10.1103/PhysRevA.92.062318}{Solving strongly
  correlated electron models on a quantum computer}, Phys. Rev. A 92 (2015)
  062318.
\newblock \href {http://dx.doi.org/10.1103/PhysRevA.92.062318}
  {\path{doi:10.1103/PhysRevA.92.062318}}.
\newline\urlprefix\url{https://link.aps.org/doi/10.1103/PhysRevA.92.062318}

\bibitem{PhysRevLett.94.170201}
M.~Troyer, U.-J. Wiese,
  \href{https://link.aps.org/doi/10.1103/PhysRevLett.94.170201}{Computational
  complexity and fundamental limitations to fermionic quantum monte carlo
  simulations}, Phys. Rev. Lett. 94 (2005) 170201.
\newblock \href {http://dx.doi.org/10.1103/PhysRevLett.94.170201}
  {\path{doi:10.1103/PhysRevLett.94.170201}}.
\newline\urlprefix\url{https://link.aps.org/doi/10.1103/PhysRevLett.94.170201}

\bibitem{Becca2017}
F.~Becca, S.~Sorella, c, 1st Edition, Cambridge University Press.
\newblock \href {http://dx.doi.org/10.1017/9781316417041}
  {\path{doi:10.1017/9781316417041}}.

\bibitem{preskill18}
J.~Preskill, \href{http://arxiv.org/pdf/1801.00862v1}{Quantum computing in the
  nisq era and beyond}, arXiv:1801.00862v1[quant-ph].
\newline\urlprefix\url{http://arxiv.org/pdf/1801.00862v1}

\bibitem{Moll2018}
N.~{Moll}, P.~{Barkoutsos}, L.~S. {Bishop}, J.~M. {Chow}, A.~{Cross}, D.~J.
  {Egger}, S.~{Filipp}, A.~{Fuhrer}, J.~M. {Gambetta}, M.~{Ganzhorn},
  A.~{Kandala}, A.~{Mezzacapo}, P.~{M{\"u}ller}, W.~{Riess}, G.~{Salis},
  J.~{Smolin}, I.~{Tavernelli}, K.~{Temme},
  \href{http://iopscience.iop.org/article/10.1088/2058-9565/aab822}{{Quantum
  optimization using variational algorithms on near-term quantum devices}},
  Quantum Science and Technology 3~(3) (2018) 030503.
\newblock \href {http://dx.doi.org/https://doi.org/10.1088/2058-9565/aab822}
  {\path{doi:https://doi.org/10.1088/2058-9565/aab822}}.
\newline\urlprefix\url{http://iopscience.iop.org/article/10.1088/2058-9565/aab822}

\bibitem{Cross2018}
A.~Cross, L.~Bishop, S.~Sheldon, N.~P.D., G.~J.M., Validating quantum computers
  using randomized model circuits, arXiv preprint arXiv:1811.12926.

\bibitem{peruzzo2014variational}
A.~Peruzzo, J.~McClean, P.~Shadbolt, M.-H. Yung, X.-Q. Zhou, P.~J. Love,
  A.~Aspuru-Guzik, J.~L. O’brien, A variational eigenvalue solver on a
  photonic quantum processor, Nature communications 5 (2014) 4213.

\bibitem{kivlichan2018quantum}
I.~D. Kivlichan, J.~McClean, N.~Wiebe, C.~Gidney, A.~Aspuru-Guzik, G.~K.-L.
  Chan, R.~Babbush, Quantum simulation of electronic structure with linear
  depth and connectivity, Physical review letters 120~(11) (2018) 110501.

\bibitem{shen_quantum_2017}
Y.~Shen, X.~Zhang, S.~Zhang, J.~Zhang, M.~Yung, K.~Kim,
  \href{http://link.aps.org/doi/10.1103/PhysRevA.95.020501}{Quantum
  implementation of the unitary coupled cluster for simulating molecular
  electronic structure}, Phys. Rev. A 95~(2) (2017) 020501.
\newblock \href {http://dx.doi.org/10.1103/PhysRevA.95.020501}
  {\path{doi:10.1103/PhysRevA.95.020501}}.
\newline\urlprefix\url{http://link.aps.org/doi/10.1103/PhysRevA.95.020501}

\bibitem{PhysRevA.92.042303}
D.~Wecker, M.~B. Hastings, M.~Troyer,
  \href{https://link.aps.org/doi/10.1103/PhysRevA.92.042303}{Progress towards
  practical quantum variational algorithms}, Phys. Rev. A 92 (2015) 042303.
\newblock \href {http://dx.doi.org/10.1103/PhysRevA.92.042303}
  {\path{doi:10.1103/PhysRevA.92.042303}}.
\newline\urlprefix\url{https://link.aps.org/doi/10.1103/PhysRevA.92.042303}

\bibitem{omalley_scalable_2016}
P.~O'Malley, R.~Babbush, I.~Kivlichan, J.~Romero, J.~McClean, R.~Barends,
  J.~Kelly, P.~Roushan, A.~Tranter, N.~Ding, B.~Campbell, Y.~Chen, Z.~Chen,
  B.~Chiaro, A.~Dunsworth, A.~Fowler, E.~J., E.~Lucero, A.~Megrant, J.~Mutus,
  M.~Neeley, C.~Neill, C.~Quintana, D.~Sank, A.~Vainsencher, J.~Wenner,
  T.~White, P.~Coveney, P.~Love, H.~Neven, A.~Aspuru-Guzik, J.~M.is,
  \href{https://link.aps.org/doi/10.1103/PhysRevX.6.031007}{Scalable {Quantum}
  {Simulation} of {Molecular} {Energies}}, Phys. Rev. X 6~(3) (2016) 031007.
\newblock \href {http://dx.doi.org/10.1103/PhysRevX.6.031007}
  {\path{doi:10.1103/PhysRevX.6.031007}}.
\newline\urlprefix\url{https://link.aps.org/doi/10.1103/PhysRevX.6.031007}

\bibitem{kandala_hardware-efficient_2017}
A.~Kandala, A.~Mezzacapo, K.~Temme, M.~Takita, M.~Brink, J.~M. Chow, J.~M.
  Gambetta,
  \href{http://www.nature.com/nature/journal/v549/n7671/full/nature23879.html?foxtrotcallback=true}{Hardware-efficient
  variational quantum eigensolver for small molecules and quantum magnets},
  Nature 549~(7671) (2017) 242--246.
\newblock \href {http://dx.doi.org/10.1038/nature23879}
  {\path{doi:10.1038/nature23879}}.
\newline\urlprefix\url{http://www.nature.com/nature/journal/v549/n7671/full/nature23879.html?foxtrotcallback=true}

\bibitem{Ganzhorn2018}
M.~Ganzhorn, D.~J. Egger, P.~K. Barkoutsos, P.~Ollitrault, G.~Salis, N.~Moll,
  A.~Fuhrer, P.~Mueller, S.~Woerner, I.~Tavernelli, S.~Filipp,
  \href{http://arxiv.org/abs/1809.050572}{Gate-efficient simulation of
  molecular eigenstates on a quantum computer}, arXiv:1809.05057.
\newline\urlprefix\url{http://arxiv.org/abs/1809.050572}

\bibitem{Hempel2018}
C.~Hempel, C.~Maier, J.~Romero, J.~McClean, T.~Monz, H.~Shen, P.~Jurcevic,
  B.~P. Lanyon, P.~Love, R.~Babbush, A.~Aspuru-Guzik, R.~Blatt, C.~F. Roos,
  \href{https://link.aps.org/doi/10.1103/PhysRevX.8.031022}{Quantum chemistry
  calculations on a trapped-ion quantum simulator}, Phys. Rev. X 8 (2018)
  031022.
\newblock \href {http://dx.doi.org/10.1103/PhysRevX.8.031022}
  {\path{doi:10.1103/PhysRevX.8.031022}}.
\newline\urlprefix\url{https://link.aps.org/doi/10.1103/PhysRevX.8.031022}

\bibitem{PhysRevLett.108.110502}
M.~Schwarz, K.~Temme, F.~Verstraete,
  \href{https://link.aps.org/doi/10.1103/PhysRevLett.108.110502}{Preparing
  projected entangled pair states on a quantum computer}, Phys. Rev. Lett. 108
  (2012) 110502.
\newblock \href {http://dx.doi.org/10.1103/PhysRevLett.108.110502}
  {\path{doi:10.1103/PhysRevLett.108.110502}}.
\newline\urlprefix\url{https://link.aps.org/doi/10.1103/PhysRevLett.108.110502}

\bibitem{supmat}
See Supplemental Material at XXXXXX for more details concerning the procedure,
  the quantum circuit for the Hubbard models, additional results, and
  benchmarks.

\bibitem{RevModPhys.66.763}
E.~Dagotto,
  \href{https://link.aps.org/doi/10.1103/RevModPhys.66.763}{Correlated
  electrons in high-temperature superconductors}, Rev. Mod. Phys. 66 (1994)
  763--840.
\newblock \href {http://dx.doi.org/10.1103/RevModPhys.66.763}
  {\path{doi:10.1103/RevModPhys.66.763}}.
\newline\urlprefix\url{https://link.aps.org/doi/10.1103/RevModPhys.66.763}

\bibitem{Dagotto257}
E.~Dagotto,
  \href{http://science.sciencemag.org/content/309/5732/257}{Complexity in
  strongly correlated electronic systems}, Science 309~(5732) (2005) 257--262.
\newblock \href
  {http://arxiv.org/abs/http://science.sciencemag.org/content/309/5732/257.full.pdf}
  {\path{arXiv:http://science.sciencemag.org/content/309/5732/257.full.pdf}},
  \href {http://dx.doi.org/10.1126/science.1107559}
  {\path{doi:10.1126/science.1107559}}.
\newline\urlprefix\url{http://science.sciencemag.org/content/309/5732/257}

\bibitem{PhysRevX.5.041041}
J.~P.~F. LeBlanc, A.~E. Antipov, F.~Becca, I.~W. Bulik, G.~K.-L. Chan, C.-M.
  Chung, Y.~Deng, M.~Ferrero, T.~M. Henderson, C.~A. Jim\'enez-Hoyos, E.~Kozik,
  X.-W. Liu, A.~J. Millis, N.~V. Prokof'ev, M.~Qin, G.~E. Scuseria, H.~Shi,
  B.~V. Svistunov, L.~F. Tocchio, I.~S. Tupitsyn, S.~R. White, S.~Zhang, B.-X.
  Zheng, Z.~Zhu, E.~Gull,
  \href{https://link.aps.org/doi/10.1103/PhysRevX.5.041041}{Solutions of the
  two-dimensional hubbard model: Benchmarks and results from a wide range of
  numerical algorithms}, Phys. Rev. X 5 (2015) 041041.
\newblock \href {http://dx.doi.org/10.1103/PhysRevX.5.041041}
  {\path{doi:10.1103/PhysRevX.5.041041}}.
\newline\urlprefix\url{https://link.aps.org/doi/10.1103/PhysRevX.5.041041}

\bibitem{Zheng1155}
B.-X. Zheng, C.-M. Chung, P.~Corboz, G.~Ehlers, M.-P. Qin, R.~M. Noack, H.~Shi,
  S.~R. White, S.~Zhang, G.~K.-L. Chan,
  \href{http://science.sciencemag.org/content/358/6367/1155}{Stripe order in
  the underdoped region of the two-dimensional hubbard model}, Science
  358~(6367) (2017) 1155--1160.
\newblock \href
  {http://arxiv.org/abs/http://science.sciencemag.org/content/358/6367/1155.full.pdf}
  {\path{arXiv:http://science.sciencemag.org/content/358/6367/1155.full.pdf}},
  \href {http://dx.doi.org/10.1126/science.aam7127}
  {\path{doi:10.1126/science.aam7127}}.
\newline\urlprefix\url{http://science.sciencemag.org/content/358/6367/1155}

\bibitem{PhysRevLett.113.046402}
P.~Corboz, T.~M. Rice, M.~Troyer,
  \href{https://link.aps.org/doi/10.1103/PhysRevLett.113.046402}{Competing
  states in the $t$-$j$ model: Uniform $d$-wave state versus stripe state},
  Phys. Rev. Lett. 113 (2014) 046402.
\newblock \href {http://dx.doi.org/10.1103/PhysRevLett.113.046402}
  {\path{doi:10.1103/PhysRevLett.113.046402}}.
\newline\urlprefix\url{https://link.aps.org/doi/10.1103/PhysRevLett.113.046402}

\bibitem{PhysRevLett.88.117002}
S.~Sorella, G.~B. Martins, F.~Becca, C.~Gazza, L.~Capriotti, A.~Parola,
  E.~Dagotto,
  \href{https://link.aps.org/doi/10.1103/PhysRevLett.88.117002}{Superconductivity
  in the two-dimensional $\mathit{t}\ensuremath{-}\mathit{J}$ model}, Phys.
  Rev. Lett. 88 (2002) 117002.
\newblock \href {http://dx.doi.org/10.1103/PhysRevLett.88.117002}
  {\path{doi:10.1103/PhysRevLett.88.117002}}.
\newline\urlprefix\url{https://link.aps.org/doi/10.1103/PhysRevLett.88.117002}

\bibitem{PhysRevB.85.081110}
W.-J. Hu, F.~Becca, S.~Sorella,
  \href{https://link.aps.org/doi/10.1103/PhysRevB.85.081110}{Absence of static
  stripes in the two-dimensional $t\ensuremath{-}j$ model determined using an
  accurate and systematic quantum monte carlo approach}, Phys. Rev. B 85 (2012)
  081110.
\newblock \href {http://dx.doi.org/10.1103/PhysRevB.85.081110}
  {\path{doi:10.1103/PhysRevB.85.081110}}.
\newline\urlprefix\url{https://link.aps.org/doi/10.1103/PhysRevB.85.081110}

\bibitem{note2}
In this case, correlated mean-field approximations suggest an uniform d-wave
  superconducting solution, while DMRG states may be biased toward a stripe
  ordering.

\bibitem{jiang2012identifying}
H.-C. Jiang, Z.~Wang, L.~Balents, Identifying topological order by entanglement
  entropy, Nature Physics 8~(12) (2012) 902.

\bibitem{PhysRevB.87.060405}
Y.~Iqbal, F.~Becca, S.~Sorella, D.~Poilblanc,
  \href{https://link.aps.org/doi/10.1103/PhysRevB.87.060405}{Gapless
  spin-liquid phase in the kagome spin-$\frac{1}{2}$ heisenberg
  antiferromagnet}, Phys. Rev. B 87 (2013) 060405.
\newblock \href {http://dx.doi.org/10.1103/PhysRevB.87.060405}
  {\path{doi:10.1103/PhysRevB.87.060405}}.
\newline\urlprefix\url{https://link.aps.org/doi/10.1103/PhysRevB.87.060405}

\bibitem{PhysRevLett.87.097201}
L.~Capriotti, F.~Becca, A.~Parola, S.~Sorella,
  \href{https://link.aps.org/doi/10.1103/PhysRevLett.87.097201}{Resonating
  valence bond wave functions for strongly frustrated spin systems}, Phys. Rev.
  Lett. 87 (2001) 097201.
\newblock \href {http://dx.doi.org/10.1103/PhysRevLett.87.097201}
  {\path{doi:10.1103/PhysRevLett.87.097201}}.
\newline\urlprefix\url{https://link.aps.org/doi/10.1103/PhysRevLett.87.097201}

\bibitem{PhysRevLett.94.026406}
M.~Capello, F.~Becca, M.~Fabrizio, S.~Sorella, E.~Tosatti,
  \href{https://link.aps.org/doi/10.1103/PhysRevLett.94.026406}{Variational
  description of mott insulators}, Phys. Rev. Lett. 94 (2005) 026406.
\newblock \href {http://dx.doi.org/10.1103/PhysRevLett.94.026406}
  {\path{doi:10.1103/PhysRevLett.94.026406}}.
\newline\urlprefix\url{https://link.aps.org/doi/10.1103/PhysRevLett.94.026406}

\bibitem{Watts1989}
J.~D. Watts, G.~W. Trucks, R.~J. Bartlett,
  \href{http://www.sciencedirect.com/science/article/pii/0009261489852467}{Coupled-cluster,
  unitary coupled-cluster and mbpt(4) open-shell analytical gradient methods},
  Chemical Physics Letters 164~(5) (1989) 502 -- 508.
\newblock \href
  {http://dx.doi.org/https://doi.org/10.1016/0009-2614(89)85246-7}
  {\path{doi:https://doi.org/10.1016/0009-2614(89)85246-7}}.
\newline\urlprefix\url{http://www.sciencedirect.com/science/article/pii/0009261489852467}

\bibitem{Bartlett1978}
R.~J. Bartlett, G.~D. Purvis,
  \href{https://onlinelibrary.wiley.com/doi/abs/10.1002/qua.560140504}{Many-body
  perturbation theory, coupled-pair many-electron theory, and the importance of
  quadruple excitations for the correlation problem}, International Journal of
  Quantum Chemistry 14~(5)  561--581.
\newblock \href
  {http://arxiv.org/abs/https://onlinelibrary.wiley.com/doi/pdf/10.1002/qua.560140504}
  {\path{arXiv:https://onlinelibrary.wiley.com/doi/pdf/10.1002/qua.560140504}},
  \href {http://dx.doi.org/10.1002/qua.560140504}
  {\path{doi:10.1002/qua.560140504}}.
\newline\urlprefix\url{https://onlinelibrary.wiley.com/doi/abs/10.1002/qua.560140504}

\bibitem{wecker2014gate}
D.~Wecker, B.~Bauer, B.~K. Clark, M.~B. Hastings, M.~Troyer, Gate-count
  estimates for performing quantum chemistry on small quantum computers,
  Physical Review A 90~(2) (2014) 022305.

\bibitem{Barkoutsos2018}
P.~K. Barkoutsos, J.~F. Gonthier, I.~Sokolov, N.~Moll, G.~Salis, A.~Fuhrer,
  M.~Ganzhorn, D.~J. Egger, M.~Troyer, A.~Mezzacapo, S.~Filipp, I.~Tavernelli,
  \href{https://link.aps.org/doi/10.1103/PhysRevA.98.022322}{Quantum algorithms
  for electronic structure calculations: Particle-hole hamiltonian and
  optimized wave-function expansions}, Phys. Rev. A 98 (2018) 022322.
\newblock \href {http://dx.doi.org/10.1103/PhysRevA.98.022322}
  {\path{doi:10.1103/PhysRevA.98.022322}}.
\newline\urlprefix\url{https://link.aps.org/doi/10.1103/PhysRevA.98.022322}

\bibitem{PhysRevA.94.032338}
J.-M. Reiner, M.~Marthaler, J.~Braum\"uller, M.~Weides, G.~Sch\"on,
  \href{https://link.aps.org/doi/10.1103/PhysRevA.94.032338}{Emulating the
  one-dimensional fermi-hubbard model by a double chain of qubits}, Phys. Rev.
  A 94 (2016) 032338.
\newblock \href {http://dx.doi.org/10.1103/PhysRevA.94.032338}
  {\path{doi:10.1103/PhysRevA.94.032338}}.
\newline\urlprefix\url{https://link.aps.org/doi/10.1103/PhysRevA.94.032338}

\bibitem{RevModPhys.63.1}
E.~Manousakis, \href{https://link.aps.org/doi/10.1103/RevModPhys.63.1}{The
  spin-1/2 heisenberg antiferromagnet on a square lattice and its application
  to the cuprous oxides}, Rev. Mod. Phys. 63 (1991) 1--62.
\newblock \href {http://dx.doi.org/10.1103/RevModPhys.63.1}
  {\path{doi:10.1103/RevModPhys.63.1}}.
\newline\urlprefix\url{https://link.aps.org/doi/10.1103/RevModPhys.63.1}

\bibitem{PhysRevB.77.024510}
L.~Spanu, M.~Lugas, F.~Becca, S.~Sorella,
  \href{https://link.aps.org/doi/10.1103/PhysRevB.77.024510}{Magnetism and
  superconductivity in the
  $t\text{\ensuremath{-}}{t}^{\ensuremath{'}}\text{\ensuremath{-}}j$ model},
  Phys. Rev. B 77 (2008) 024510.
\newblock \href {http://dx.doi.org/10.1103/PhysRevB.77.024510}
  {\path{doi:10.1103/PhysRevB.77.024510}}.
\newline\urlprefix\url{https://link.aps.org/doi/10.1103/PhysRevB.77.024510}

\bibitem{Jastrow}
R.~Jastrow, \href{https://link.aps.org/doi/10.1103/PhysRev.98.1479}{Many-body
  problem with strong forces}, Phys. Rev. 98 (1955) 1479--1484.
\newblock \href {http://dx.doi.org/10.1103/PhysRev.98.1479}
  {\path{doi:10.1103/PhysRev.98.1479}}.
\newline\urlprefix\url{https://link.aps.org/doi/10.1103/PhysRev.98.1479}

\bibitem{casula2004correlated}
M.~Casula, C.~Attaccalite, S.~Sorella, Correlated geminal wave function for
  molecules: An efficient resonating valence bond approach, The Journal of
  chemical physics 121~(15) (2004) 7110--7126.

\bibitem{zen2014static}
A.~Zen, E.~Coccia, Y.~Luo, S.~Sorella, L.~Guidoni, Static and dynamical
  correlation in diradical molecules by quantum monte carlo using the jastrow
  antisymmetrized geminal power ansatz, Journal of chemical theory and
  computation 10~(3) (2014) 1048--1061.

\bibitem{PhysRevLett.120.025701}
G.~Mazzola, R.~Helled, S.~Sorella,
  \href{https://link.aps.org/doi/10.1103/PhysRevLett.120.025701}{Phase diagram
  of hydrogen and a hydrogen-helium mixture at planetary conditions by quantum
  monte carlo simulations}, Phys. Rev. Lett. 120 (2018) 025701.
\newblock \href {http://dx.doi.org/10.1103/PhysRevLett.120.025701}
  {\path{doi:10.1103/PhysRevLett.120.025701}}.
\newline\urlprefix\url{https://link.aps.org/doi/10.1103/PhysRevLett.120.025701}

\bibitem{genovese}
C.~Genovese, A.~Meninno, S.~Sorella,
  \href{https://doi.org/10.1063/1.5081933}{Assessing the accuracy of the
  jastrow antisymmetrized geminal power in the h4 model system}, The Journal of
  Chemical Physics 150~(8) (2019) 084102.
\newblock \href {http://arxiv.org/abs/https://doi.org/10.1063/1.5081933}
  {\path{arXiv:https://doi.org/10.1063/1.5081933}}, \href
  {http://dx.doi.org/10.1063/1.5081933} {\path{doi:10.1063/1.5081933}}.
\newline\urlprefix\url{https://doi.org/10.1063/1.5081933}

\bibitem{neu1}
E.~Neuscamman, \href{https://doi.org/10.1063/1.4829835}{The jastrow
  antisymmetric geminal power in hilbert space: Theory, benchmarking, and
  application to a novel transition state}, The Journal of Chemical Physics
  139~(19) (2013) 194105.
\newblock \href {http://arxiv.org/abs/https://doi.org/10.1063/1.4829835}
  {\path{arXiv:https://doi.org/10.1063/1.4829835}}, \href
  {http://dx.doi.org/10.1063/1.4829835} {\path{doi:10.1063/1.4829835}}.
\newline\urlprefix\url{https://doi.org/10.1063/1.4829835}

\bibitem{neu2}
E.~Neuscamman, \href{https://doi.org/10.1063/1.4829536}{Communication: A
  jastrow factor coupled cluster theory for weak and strong electron
  correlation}, The Journal of Chemical Physics 139~(18) (2013) 181101.
\newblock \href {http://arxiv.org/abs/https://doi.org/10.1063/1.4829536}
  {\path{arXiv:https://doi.org/10.1063/1.4829536}}, \href
  {http://dx.doi.org/10.1063/1.4829536} {\path{doi:10.1063/1.4829536}}.
\newline\urlprefix\url{https://doi.org/10.1063/1.4829536}

\bibitem{PhysRevLett.10.159}
M.~C. Gutzwiller,
  \href{https://link.aps.org/doi/10.1103/PhysRevLett.10.159}{Effect of
  correlation on the ferromagnetism of transition metals}, Phys. Rev. Lett. 10
  (1963) 159--162.
\newblock \href {http://dx.doi.org/10.1103/PhysRevLett.10.159}
  {\path{doi:10.1103/PhysRevLett.10.159}}.
\newline\urlprefix\url{https://link.aps.org/doi/10.1103/PhysRevLett.10.159}

\bibitem{RevModPhys.73.33}
W.~M.~C. Foulkes, L.~Mitas, R.~J. Needs, G.~Rajagopal,
  \href{https://link.aps.org/doi/10.1103/RevModPhys.73.33}{Quantum monte carlo
  simulations of solids}, Rev. Mod. Phys. 73 (2001) 33--83.
\newblock \href {http://dx.doi.org/10.1103/RevModPhys.73.33}
  {\path{doi:10.1103/RevModPhys.73.33}}.
\newline\urlprefix\url{https://link.aps.org/doi/10.1103/RevModPhys.73.33}

\bibitem{PhysRevLett.120.110501}
I.~D. Kivlichan, J.~McClean, N.~Wiebe, C.~Gidney, A.~Aspuru-Guzik, G.~K.-L.
  Chan, R.~Babbush,
  \href{https://link.aps.org/doi/10.1103/PhysRevLett.120.110501}{Quantum
  simulation of electronic structure with linear depth and connectivity}, Phys.
  Rev. Lett. 120 (2018) 110501.
\newblock \href {http://dx.doi.org/10.1103/PhysRevLett.120.110501}
  {\path{doi:10.1103/PhysRevLett.120.110501}}.
\newline\urlprefix\url{https://link.aps.org/doi/10.1103/PhysRevLett.120.110501}

\bibitem{verstraete2005mapping}
F.~Verstraete, J.~I. Cirac, Mapping local hamiltonians of fermions to local
  hamiltonians of spins, Journal of Statistical Mechanics: Theory and
  Experiment 2005~(09) (2005) P09012.

\bibitem{Qiskit}
G.~Aleksandrowicz, T.~Alexander, P.~Barkoutsos, L.~Bello, Y.~Ben-Haim,
  D.~Bucher, F.~J. Cabrera-Hern{\'a}dez, J.~Carballo-Franquis, A.~Chen, C.-F.
  Chen, J.~M. Chow, A.~D. C{\'o}rcoles-Gonzales, A.~J. Cross, A.~Cross,
  J.~Cruz-Benito, C.~Culver, S.~D. L.~P. Gonz{\'a}lez, E.~D.~L. Torre, D.~Ding,
  E.~Dumitrescu, I.~Duran, P.~Eendebak, M.~Everitt, I.~F. Sertage, A.~Frisch,
  A.~Fuhrer, J.~Gambetta, B.~G. Gago, J.~Gomez-Mosquera, D.~Greenberg,
  I.~Hamamura, V.~Havlicek, J.~Hellmers, {\L}.~Herok, H.~Horii, S.~Hu,
  T.~Imamichi, T.~Itoko, A.~Javadi-Abhari, N.~Kanazawa, A.~Karazeev,
  K.~Krsulich, P.~Liu, Y.~Luh, Y.~Maeng, M.~Marques, F.~J.
  Mart{\'\i}n-Fern{\'a}ndez, D.~T. McClure, D.~McKay, S.~Meesala, A.~Mezzacapo,
  N.~Moll, D.~M. Rodr{\'\i}guez, G.~Nannicini, P.~Nation, P.~Ollitrault, L.~J.
  O'Riordan, H.~Paik, J.~P{\'e}rez, A.~Phan, M.~Pistoia, V.~Prutyanov,
  M.~Reuter, J.~Rice, A.~R. Davila, R.~H.~P. Rudy, M.~Ryu, N.~Sathaye,
  C.~Schnabel, E.~Schoute, K.~Setia, Y.~Shi, A.~Silva, Y.~Siraichi,
  S.~Sivarajah, J.~A. Smolin, M.~Soeken, H.~Takahashi, I.~Tavernelli,
  C.~Taylor, P.~Taylour, K.~Trabing, M.~Treinish, W.~Turner, D.~Vogt-Lee,
  C.~Vuillot, J.~A. Wildstrom, J.~Wilson, E.~Winston, C.~Wood, S.~Wood,
  S.~W{\"o}rner, I.~Y. Akhalwaya, C.~Zoufal, Qiskit: An open-source framework
  for quantum computing (2019).
\newblock \href {http://dx.doi.org/10.5281/zenodo.2562110}
  {\path{doi:10.5281/zenodo.2562110}}.

\bibitem{note3}
For example, we have $\langle XI + IX \rangle = -2~ x_{00}+0 ~x_{10}+0~
  x_{01}+2~ x_{11}$,where $x_i$ is the normalized count of the 2-bit string
  readout $i$ in basis $\{ X,X\}$.

\bibitem{motta2019quantum}
M.~Motta, C.~Sun, A.~T.~K. Tan, M.~J. Rourke, E.~Ye, A.~J. Minnich, F.~G.
  Brandao, G.~K. Chan, Quantum imaginary time evolution, quantum lanczos, and
  quantum thermal averaging, arXiv preprint arXiv:1901.07653.

\bibitem{mcardle2018error}
S.~McArdle, X.~Yuan, S.~Benjamin,
  \href{https://link.aps.org/doi/10.1103/PhysRevLett.122.180501}{Error-mitigated
  digital quantum simulation}, Phys. Rev. Lett. 122 (2019) 180501.
\newblock \href {http://dx.doi.org/10.1103/PhysRevLett.122.180501}
  {\path{doi:10.1103/PhysRevLett.122.180501}}.
\newline\urlprefix\url{https://link.aps.org/doi/10.1103/PhysRevLett.122.180501}

\end{thebibliography}

%%%%%%%%%% Merge with supplemental materials %%%%%%%%%%
\pagebreak
\widetext
\begin{center}
\textbf{\large Supplemental Materials: Non-unitary operations for ground-state calculations in near term quantum computers}
\end{center}
%%%%%%%%%% Merge with supplemental materials %%%%%%%%%%
%%%%%%%%%% Prefix a "S" to all equations, figures, tables and reset the counter %%%%%%%%%%
\setcounter{equation}{0}
\setcounter{figure}{0}
\setcounter{table}{0}
\setcounter{page}{1}
%\makeatletter
\renewcommand{\theequation}{S\arabic{equation}}
\renewcommand{\thefigure}{S\arabic{figure}}
\renewcommand{\bibnumfmt}[1]{[S#1]}
\renewcommand{\citenumfont}[1]{S#1}
%%%%%%%%%% Prefix a "S" to all equations, figures, tables and reset the counter %%%%%%%%%%

\section{An illustrative example}

\begin{figure}[hbt]
\includegraphics[width=0.8\columnwidth]{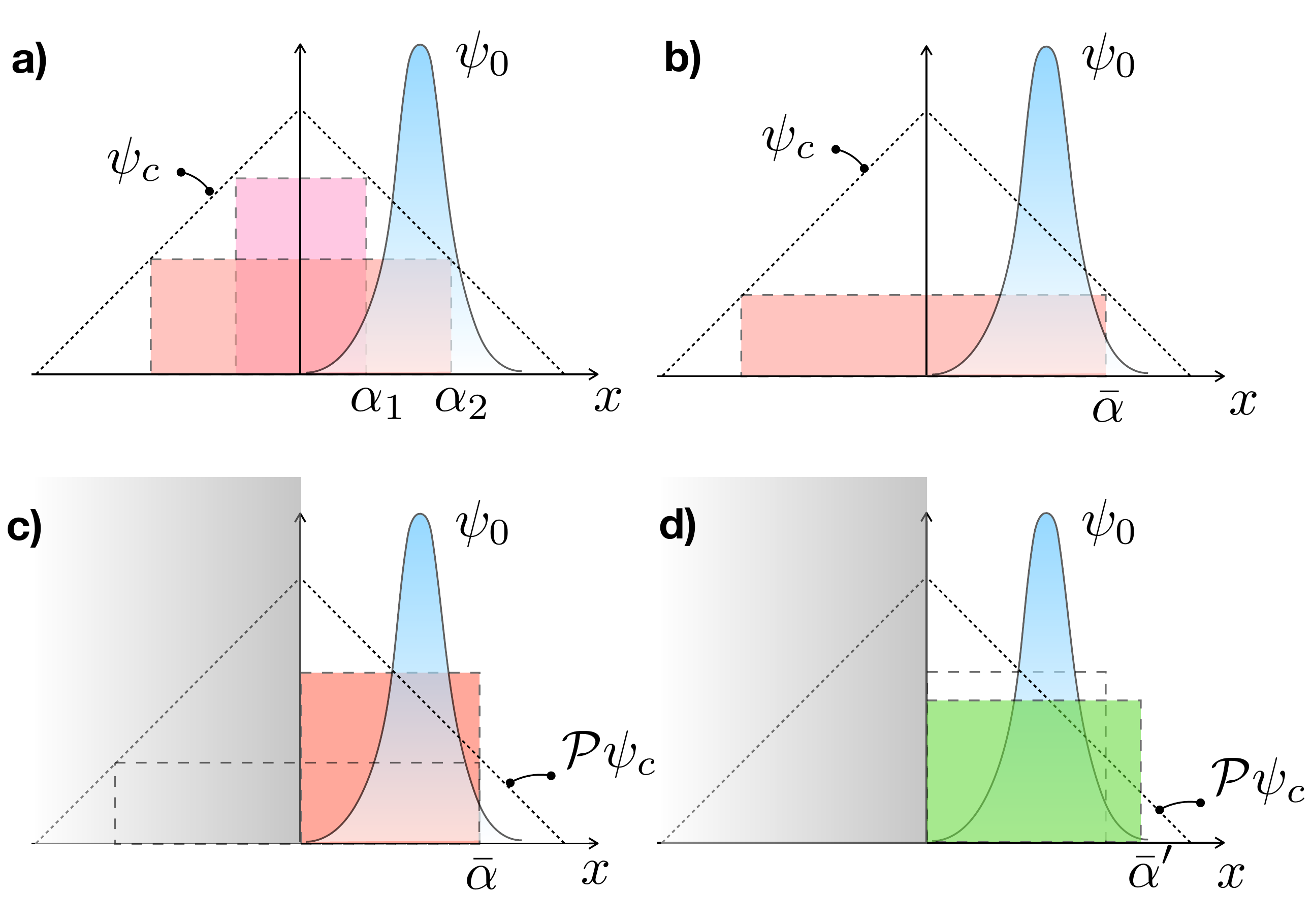}\caption{
\emph{Illustrative example of projected variational states}. Panel (a): The target state $\psi_0$ (blue), exact solution of the Hamiltonian of the problem, and two instances (pink) of  rectangle-shaped variational states $\psi_c(\alpha_1)$ and $\psi_c(\alpha_2)$. (b) Variational state $\psi_c(\bar\alpha)$, having best overlap with the target state. (c) A simple projector $\mathcal{P}$ on the $x>0$ plane filters out most of the components of $\psi_c$ (gray area) having vanishing overlap with $\psi_0$. Upon renormalization this produces the class of states $\mathcal{P} \psi_c$. (d) The optimal circuit parameter $\bar\alpha'$, in the presence of  $\mathcal{P}$ may be generally different from $\bar\alpha$. The new optimal solution (green) has much larger overlap with the target state compared with the simple circuit ansatz of panel (b).
}
\label{fig:ske}
\end{figure}

 In Fig.~\ref{fig:ske} we sketch a typical situation where the target state, being the exact ground state of a onedimensional model, $\psi_0(x)$ is outside the trial state functional class, $\psi_c(x)$.
In this example the variational freedom is quite small, limiting the maximum overlap $\langle \psi_c | \psi_0 \rangle$ available.
We can improve the ansatz quality by introducing a suitable projector $\mathcal{P}$, which removes the components of $|\psi_c\rangle$  having negligible overlap. In principle the functional form of projector operator is in general defined through variational parameters, which must be optimized together with the ones contained in the circuit.

\clearpage
\section{Size scaling of the computational gain}

We observe that the computational gain remains meaningful (i.e. order-of-magnitutes large) as the system size $L$ is increased. We measure this gain, at fixed circuit depth $d$, as the ratio of the energy differences (compared to the exact value $E_{exact}$) between the best energy obtained with the circuit, $E_c$, and the best energy obtained by the JQC ansatz, $E_{JQC}$.
These optimizations are independent.
We notice (not shown) that the circuit parameters $\bm{\theta}$ optimized in the presence of the Jastrow become sub-optimal if plugged back into the circuit state alone. This observation is common in the context of classical QMC calculations, where it is common practice to optimize together the  parameters defining the Jastrow and the Slater-Determinant.

The numerical optimizations of the variational forms are done running first the COBYLA optimizer, followed by a BFGS run.

The computational gain is model dependent, and   it becomes sizable only for $d>1$ for the Ising model.
In this case, the variational freedom provided by the (short) circuit state $| \psi_c\rangle$ is not sufficiently large and the Jastrow projector acts only trivially on the state.
For all system sizes, the computational gain is increasing with increasing depth. 

\begin{figure}[hbt]
\includegraphics[width=1.0\columnwidth]{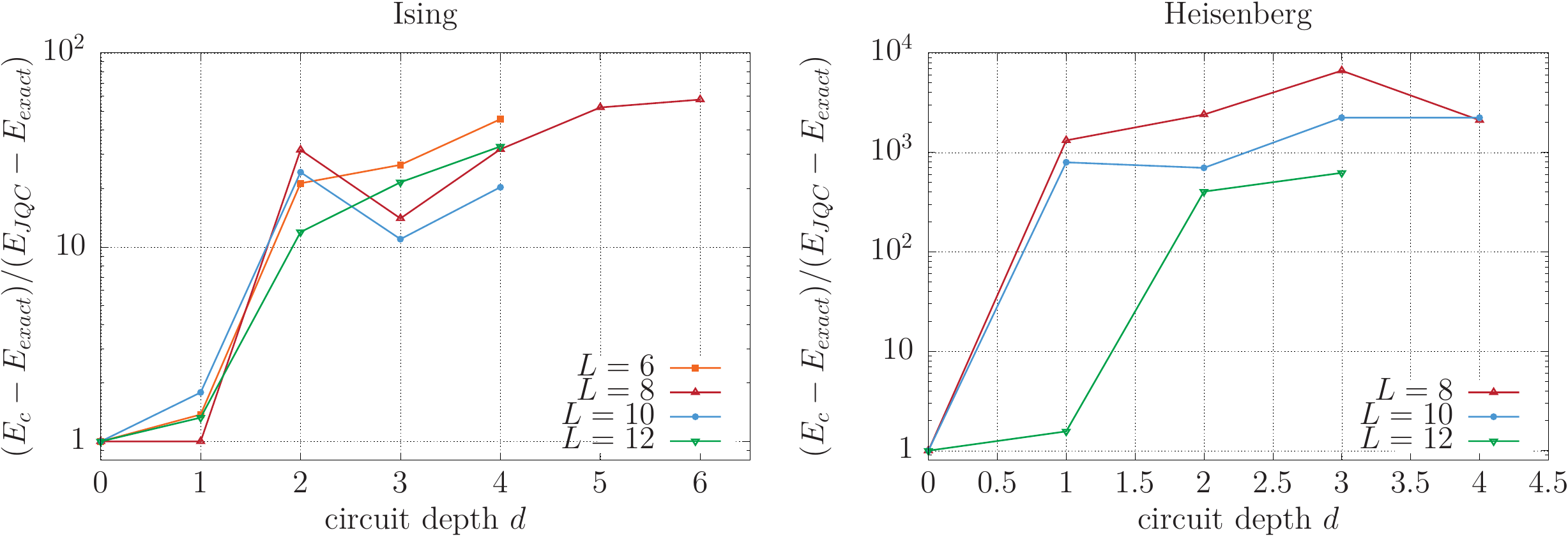}\caption{
Computational gain provided by the Jastrow operator as a function of the circuit depth. This is an $R_y$-CNOT heuristic circuit (cfn. Fig.~\ref{fig:rycnot}).
}
\label{fig:size}
\end{figure}

It is not possible to obtain a meaningful size-scaling in the case of the Hubbard model, since the Jordan-Wigner mapping already doubles the  qubits requirement to encode the system.
While the $L=2$ system is so simple that already a $d=1$ $R_y$CNOT is sufficient to recover the exact energy, obtaining a set of converged numerical optimizations up to the $L=8$ point (hence $N=16$) is too computationally expensive (for classical emulation).

\begin{figure}[t]
	\begin{tikzpicture}
    	\draw[dashed] (-2.5,-2) -- (1.5,-2);
        \draw[dashed] (-2.5,-2) -- (-2.5,2);
        \draw[dashed] (1.5,2.2) -- (1.5,-1.8);
        \draw[dashed] (1.5,2.2) -- (-2,2.2);
               
        %\draw[dashed] (2.15,-1.8) -- (3.5,-1.8);
        %\draw[dashed] (3.5, -1.8) -- (3.5, 2);
        %\draw[dashed] (2.15,-1.8) -- (2.15,2);
        %\draw[dashed] (2.15, 2) -- (3.5, 2);
        
        \node at (0.0,2.5) { $d$-times};
        
		\node at (0,0) {
		\Qcircuit @C=0.93em @R=.23em {
		&q_{N-1} & \quad & \quad   & \ket{0} & \quad & \qw & \qw & \gate{ \ R(\vec{\theta}_{q_N}) \ \ }  & \multigate{4}{U_{\rm{ENT}}} 	& \qw & \qw & \qw &\gate{ \ \mathcal{R}^{N \ \ }_{\alpha}} & \meter\\
		&q_{N-2}    & \quad & \quad   & \ket{0} &\quad & \qw & \qw  & \gate{ R(\vec{\theta}_{q_N-1})} & \ghost{U_{\rm{ENT}}}	& \qw & \qw & \qw &\gate{\mathcal{R}^{N-1}_{\alpha}} & \meter\\
		& \dots   & \quad & \quad &  \ket{0} & \quad & \qw & \qw & \gate{ \ R(\vec{\theta}_{q_{\dots}}) \ \ } & \ghost{U_{\rm{ENT}}}	&\qw & \qw & \qw &\gate{ \ \mathcal{R}^{\dots \ \ }_{\alpha}} & \meter\\
		&q_1    & \quad & \quad   & \ket{0} & \quad & \qw & \qw & \gate{ \ R(\vec{\theta}_{q_{1}}) \ \ \ } & \ghost{U_{\rm{ENT}}}	&\qw & \qw & \qw &\gate{ \ \mathcal{R}^{1 \ \ }_{\alpha}} & \meter\\
		&q_0 	    & \quad & \quad   & \ket{0} &\quad & \qw & \qw & \gate{ \ R(\vec{\theta}_{q_{0}}) \ \ \ } & \ghost{U_{\rm{ENT}}} & \qw & \qw & \qw &\gate{ \ \mathcal{R}^{0 \ \ }_{\alpha}} & \meter \\ }
		};
	\end{tikzpicture}
	\caption{Heuristic circuit used in this work. The circuit is made of $d$ repetitions of the same type of block, which features single qubit parametrized rotations on the $y$ axis, and an entangler block made of a cascade of CNOT gates (cfn. also Fig.~\ref{fig:hubbb}.d).
	}
    \label{fig:rycnot}
\end{figure}
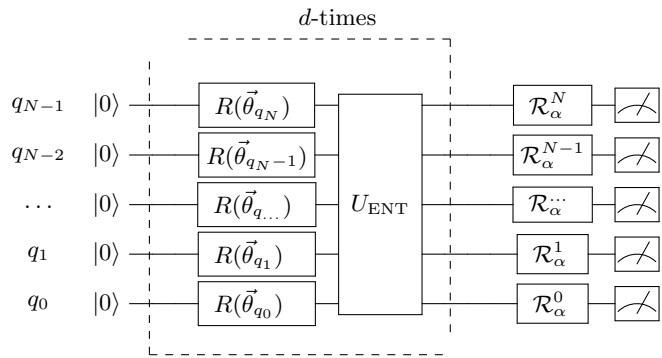

\section{Hubbard model in a quantum computer}

\subsection{Fermions-to-qubits mapping}

We consider the standard mapping of the $L$ sites Hubbard chain with hamiltonian
\begin{equation}
\label{eq:hub2}
    \mathcal{H}_{Hub} = -t \sum_{i=0}^{L-1} \sum_{s=\uparrow,\downarrow} (c_{i,s}^{\dag} c_{i+1,s} + c_{i+1,s}^{\dag} c_{i,s}) +
    U \sum_{i=0}^{L-1} (c_{i,\uparrow}^{\dag} c_{i,\uparrow}  c_{i,\downarrow}^{\dag} c_{i,\downarrow}) - \frac U 2 \sum_{s=\uparrow,\downarrow} \sum_{i=0}^{L-1} c_{i,s}^{\dag} c_{i,s} + \frac U 4 ,
\end{equation}
where we tuned the chemical potential to satisfy the half-filling condition,
to a $2L$ qubits ladder system having the following hamiltonian:
\begin{equation}
\label{eq:hub2}
    \mathcal{H}_{Hub}^Q = - \frac t 2 \sum_{i=0}^{L-1} \sum_{s=\uparrow,\downarrow} (\sigma^x_{i,s} \sigma^x_{i+1,s} + \sigma^y_{i,s} \sigma^y_{i+1,s} )+
    \frac U 4 \sum_{i=0}^{L-1} (\sigma^z_{i,\uparrow}+1)(\sigma^z_{i,\downarrow}+1) - \frac U 4 \sum_{i=0}^{L-1}\sum_{s=\uparrow,\downarrow} \sigma^z_{i,s},
\end{equation}
where now the $s=\uparrow(\downarrow)$ qubits correspond to the lower(upper) chain of the ladder (see Fig.~\ref{fig:hubbb}.a).
If two vertically adjacent qubits (red link) are in the state $1$, they contribute with a Coulomb repulsion term $+U$ to the total energy.

\begin{figure}[hbt]
\includegraphics[width=1.0\columnwidth]{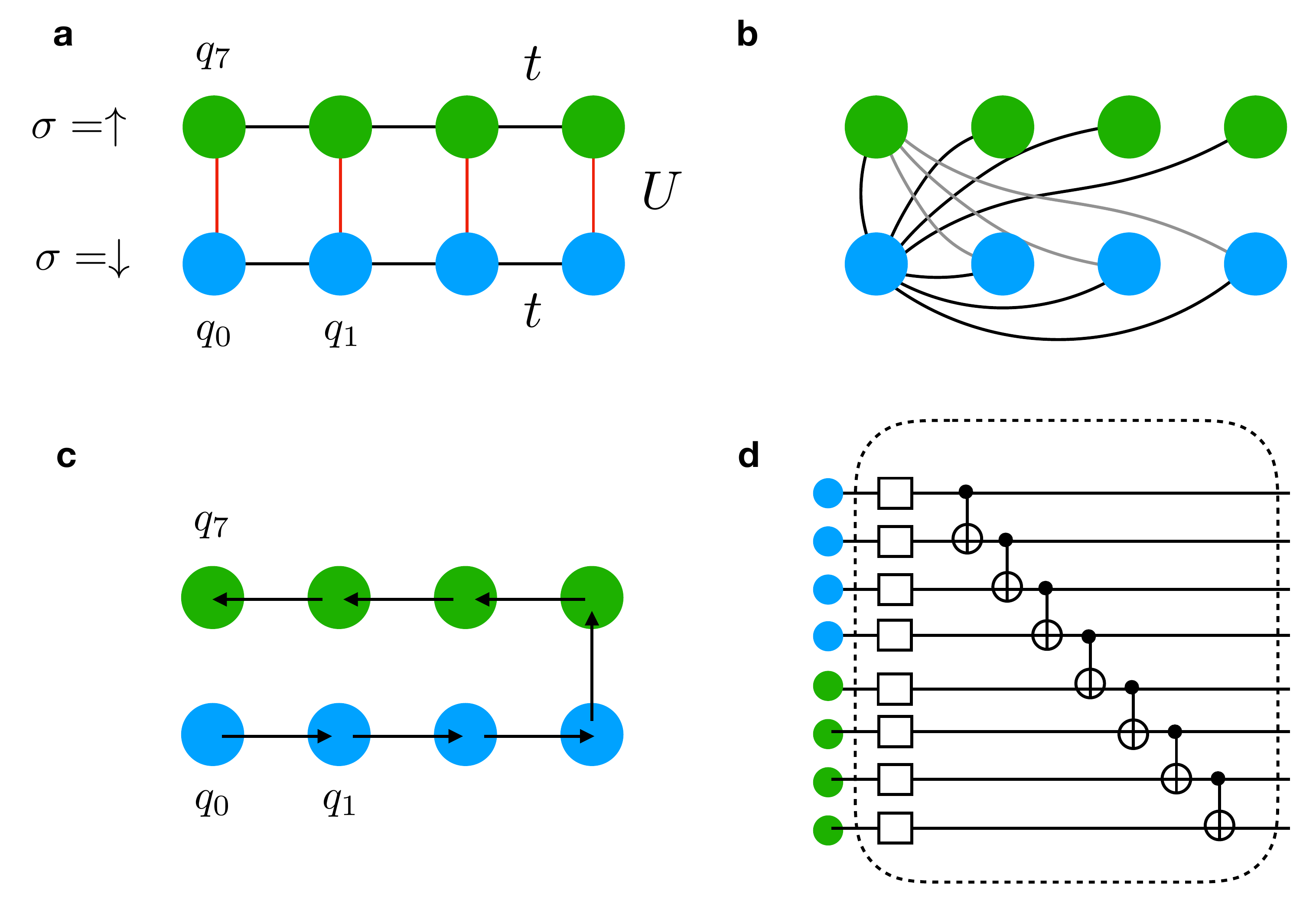}\caption{
One-dimensional Hubbard model in a quantum computer. a) The structure of the equivalent qubits ladder chain. b) The inequivalent qubit-qubit interaction considered in the Jastrow operator. c) The specific qubit enumeration used in this work. d) The structure of the entangler block of the heuristic circuit. Squares represent $R_y$ rotation single qubit gates. Each rotation introduces a variational parameter. Qubit $q_i$ controls the target  qubit $q_{i+1}$ in the CNOT cascade.
}
\label{fig:hubbb}
\end{figure}

\subsection{Symmetries in the Jastrow operator}
While the total number of parameters in a two-spin Jastrow correlator is $N(N-1)/2$, this number can be reduced according the system's symmetries. This parameter reduction alleviates the local minima problem during optimization.

In this case we reduce the number of optimizable parameters from a total of 28 to 10 inequivalent spin-spin interaction. Notice that, since the Jastrow operator is not applied at the circuit level, the available interactions are not constrained by the hardware connectivity.

Due to translational invariance we assume that the $q_0$-$q_1$, $q_1$-$q_2$, etc. interactions are equal, so that the $\lambda_{01}, \lambda_{12}$, etc. parameters are also equal (see main text). The same applies for all the next-nearest-neighbours $q_0$-$q_2$, $q_1$-$q_3$, etc. along the same chain, and so on.
In Fig.~\ref{fig:hubbb}.b we display all the inequivalent coupling parameters.

\subsection{Heuristic circuit}

The qubits are enumerated as in Fig.~\ref{fig:hubbb}.c. Different implementations are also possible.
The structure of one of the entangler blocks of the heuristic circuit is  shown also in Fig.~\ref{fig:hubbb}.d.
It is not the purpose of the paper to optimize the circuit architecture for this specific system.

\section{Probabilities reconstruction in the entangled copy method}

The \emph{implementation \#A} (as defined in the main text) uses an ancilla register. 

The total register now reads $\{q_0,\cdots,q_{L-1}.q_{L},\cdots, q_{2L-1} \}$, where the first $L$ qubits evolve through the circuit, and the last $L$ ones are initialized to 0. 
We use the ancilla register to store an entangled copy of the original register, using CNOT gates as in described in the main text.
%Each ancillary qubits $i_a$ is controlled by the corresponding circuit qubit $i_c = i_a-N$ via a CNOT gate.
If we consider the $L=2$ Ising model and the $H \otimes H$ circuit (for sake of simplicity) outlined in the main text,
before applying the post-rotations, the circuit wavefunction encoded in the total register  reads
$\psi_c= 1/2 ( |00.00\rangle +  |01.01\rangle +  |10.10\rangle + |11.11\rangle )$.
The first two qubits, i.e. the circuit (or system) qubits, are then rotated accordingly to the desired Hamiltonian term to be measured, whereas the ancillary qubit are read in their Z basis.

Let us call $|i\rangle$ an $L$ bit string encoded in the system register and $|j\rangle$ the $L$ bit string encoded in the ancilla register, with $i,j = 1, \cdots, 2^L-1$. The state $|j.i\rangle$, belonging to the $2^{2L}$ dimensional Hilbert space of the total register, encodes in binary format the number $|j.i\rangle \rightarrow j ~2^L + i$.
Our goal is to reconstruct the probability of the system register in any possible basis, which is obtained by applying the postrotations.

Applying $L$ postrotation $\mathcal{R_\alpha}$, with $\alpha=[Z,X,Y]$, to the system register, is mathematically translated into applying an unitary operator,  tensor product of the individual gates,
\begin{equation}
    \mathcal{U}_{\bf b} = \mathcal{R}_{\alpha_0} \otimes \mathcal{R}_{\alpha_1} \cdots  \otimes \mathcal{R}_{\alpha_L},
\end{equation}
where the basis ${\bf b}$, is univocally defined by the string $[{\alpha_0},{\alpha_0}, \cdots, {\alpha_L}] $,
and the three possible post-rotation unitaries are defined as
%\begin{equation}
%    \mathcal{R}_I = 
%  \left[ {\begin{array}{cc}
%   1 & 2 \\
%   3 & 4 \\
%  \end{array} } \right]
%\end{equation}
\begin{align}
\mathcal{R}_Z &= \begin{bmatrix}
    1 & ~0 \\
    0 & ~1
\end{bmatrix},
&
\mathcal{R}_X &= {\frac{1}{ \sqrt 2}} \begin{bmatrix}
    1 & ~1 \\
    1 & -1
\end{bmatrix},
&
\mathcal{R}_Y &= {\frac{1}{ \sqrt 2}} \begin{bmatrix}
    1 & -i \\
    1 & ~i
\end{bmatrix}.
\end{align}
To give an example, for the Ising model we will use the basis ${\bf b}_0 = [Z,Z, \cdots, Z]$, and ${\bf b}_1 = [X,X, \cdots, X]$.
In general we will have a $B$ number of basis.

Recall that we need to reconstruct $B$ $L$-qubit probabilities to follow the present way to calculate the expectation values of the energy, written as a sum of Pauli operators, as described in the main text. In general, this may not be the most efficient way to achieve this task, but it is the widely adopted one.
In the Ising model example however, the total energy is computed evaluating the (two-body) operator
$ Z \otimes Z \otimes I \cdots I + I \otimes Z \otimes Z \cdots \otimes I + \cdots + I \otimes \cdots \otimes Z \otimes Z $ on the probability $P_{{\bf b}0}$, and the (one-body) operator $ Z \otimes I \otimes I \cdots I + I \otimes Z \otimes I \cdots \otimes I + \cdots + I \otimes \cdots \otimes I \otimes Z $ on the probability $P_{{\bf b}1}$, where all the system qubits are measured along the X axis.

\subsection{Positive valued states.}
Reconstructing all the $B$ $L-$qubits probabilities arrays is far from being trivial, except for the ${\bf b}_0 = [Z,Z, \cdots, Z]$ basis case.
Using the entangled copy method we can directly measure $2L-$qubits probabilities arrays instead.
For the sake of illustration, we provide here a simple method that works only when the quantum state we are sampling from is positive-valued. 
Since the Ising model hardware calculations fall within this class we provide here the specific formula used, and discuss the generalization to the non-positive case in the next section.

In this case, 
to extract the $L-$qubits probability $P_{\bf b}$ , given the measured  $2L-$qubits probability $\bar P_{\bf b}$, in the basis ${\bf b}$, we use the following formula,
\begin{equation}
    P_{\bf b}(i) =\left[ \sum_{j=0}^{2^L-1} \Lambda_{\bf b}(i,j) ~\sqrt{\bar P_{\bf b}(j ~2^L + i) } \right]^2,
\label{eq:rec}
\end{equation}
where $\Lambda_{\bf b}(i,j) = 2^{m/2} ~\mathcal{U}_{\bf b}(i,j) $, with $m=L-\#$of $\mathcal{R}_I$ rotations in the tensor $\mathcal{U}_{\bf b}$, and $ [~ ]^2$ implies taking the square modulus of the complex number.

For the sake of concreteness, in the case of the L-site Ising model, the two matrices used are the tensor product of L identity (hadamard) matrices for ${\bf b}_0$ (${\bf b}_1$). E.g. for $L=2$,
\begin{align}
  \Lambda_{{\bf b}0} &= \begin{bmatrix}
    1 & ~ 0 &  ~0 &  ~0\\
    0 & ~1 &  ~0 & ~0\\
    0 & ~ 0 & ~1 & ~0\\
    0 & ~0 & ~0 &  ~1
  \end{bmatrix},
&
  \Lambda_{{\bf b}1} &= \begin{bmatrix}
    1 &  ~1 &  ~1 &  ~1\\
    1 & -1 &  ~1 & -1\\
    1 &  ~1 & -1 & -1\\
    1 & -1 & -1 &  ~1
  \end{bmatrix}.
\label{eq:case}
\end{align}

Notice that, while formally the sum extends to an exponentially increasing number of components, it is in practice restricted to the number of non-zero $\bar P_{\bf b}(j ~2^L + i)$ terms measured, which is at most equal to the maximum number of measurement performed.
Bounding the number of measurements required to achieve a target energy accuracy is a central problem in VQE, and its outside the scope of this work.

\begin{figure}[hbt]
\includegraphics[width=1\columnwidth]{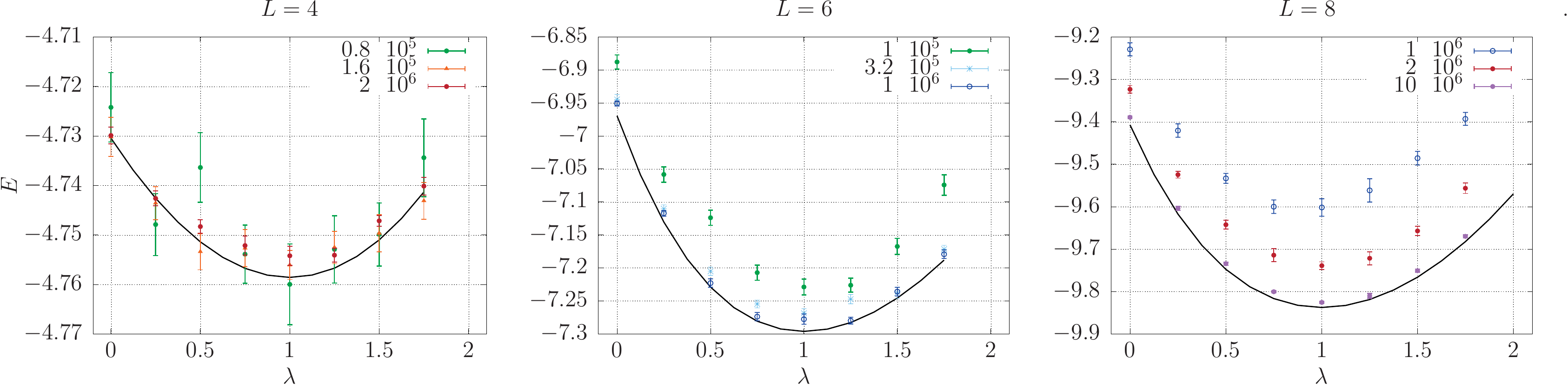}\caption{
Energy of the JQC states for the Ising model with different size as a function of the reparametrized Jastrow coefficients.
The $\lambda$ value  controls an element-wise re-parametrization of the optimal Jastrow parameters, such that when $\lambda=0$ the JQC state reduces to the circuit state $\psi_c$ and no measurement reweighting is applied, whereas for  $\lambda=1$ the optimal Jastrow parameters are recovered and the JQC state has the minimum energy.
In black we plot the numerically obtained value via state-vector evaluation of the JQC ansatz.
Colored are the noiseless circuit simulation values, that include the extended circuit used for the entangled copy implememntation and the reweighting procedure.
Different colors represent  different measurement numbers used to compute the probabilities $\bar P_{\bf b}$ in the two basis considered for the Ising model.
Error bars are  computed by repeating the process $M_{rep}=12$ times. 
}
\label{fig:sssize}
\end{figure}

Overall, the number of measurements determines the quality of the result, as shown in Fig.~\ref{fig:sssize}, where we study the accuracy of the energy  compared with the theoretical value provided by the JQC ansatz.
An imperfect reconstruction of the probabilities leads to a systematically larger value for the energy.

\clearpage

\subsection{Non-positive states.}

Eq.~\ref{eq:rec} does not work in the case of non-positive wavefunction. The knowledge of only $\bar P_{\bf b}$ is not sufficent to obtain $ P_{\bf b}$ because the information about the signs of the components is erased when computing the square of the amplitutes.
Our goal is however feasible because we can always obtain the missing information in an independent way. This is possible if we do not apply the entangled copy circuit (notice however that without the entangled copy circuit we cannot apply to Jastrow projector).
Conceptually we need to optimize the $\Lambda_{\bf b}$ matrix of Eq.~\ref{eq:rec} in such a way that, given the  $\bar P_{\bf b}$, the reduced probability computed via the postprocessing,  $ P_{\bf b}$, is equal -or close- to the same probability distribution $ P^0_{\bf b}$ measured from the same state $\psi_c$ without the entangled copy circuit appended.

In practice we find that imposing a particular structure for the matrix $\Lambda_{\bf b}$ is beneficial.
We inherit such structure from the "positive case", but we multiply element-wise each column $k$ by a number $s_k$ which should ideally take discrete values $\pm 1$. For example, $ \Lambda_{{\bf b}1}$ in Eq~\ref{eq:case} becomes 
\begin{align}
   \Lambda_{{\bf b}1} &= \begin{bmatrix}
    s_0 &  ~s_1 &  ~s_2 &  ~s_3\\
    s_0 & -s_1 &  ~s_2 & -s_3\\
    s_0 &  ~s_1 & -s_2 & -s_3\\
    s_0 & -s_1 & -s_2 &  ~s_3
  \end{bmatrix}.
\label{eq:modif}
\end{align}
We then need to solve numerically an optimization problem, involving a system of quadratic equations, minimizing the difference
\begin{equation}
\label{eq:opti}
    \min_{\bf s} | P_{\bf b} -  P^0_{\bf b} |
\end{equation}
where $\bf s$ is the solution  array $\{s_0, s_1, \cdots\}$, where the probability array $P_{\bf b}$ is obtained inserting Eq.~\ref{eq:modif} into Eq.~\ref{eq:case}, while $P^0_{\bf b}$ can be simply measured without the entangled copy circuit.

It is important to observe that we require only an approximate solution of this system, as the reweighting procedure that will follows will anyway compensate for possible errors in the reconstruction.

\begin{figure}[hbt]
\includegraphics[width=0.5\columnwidth]{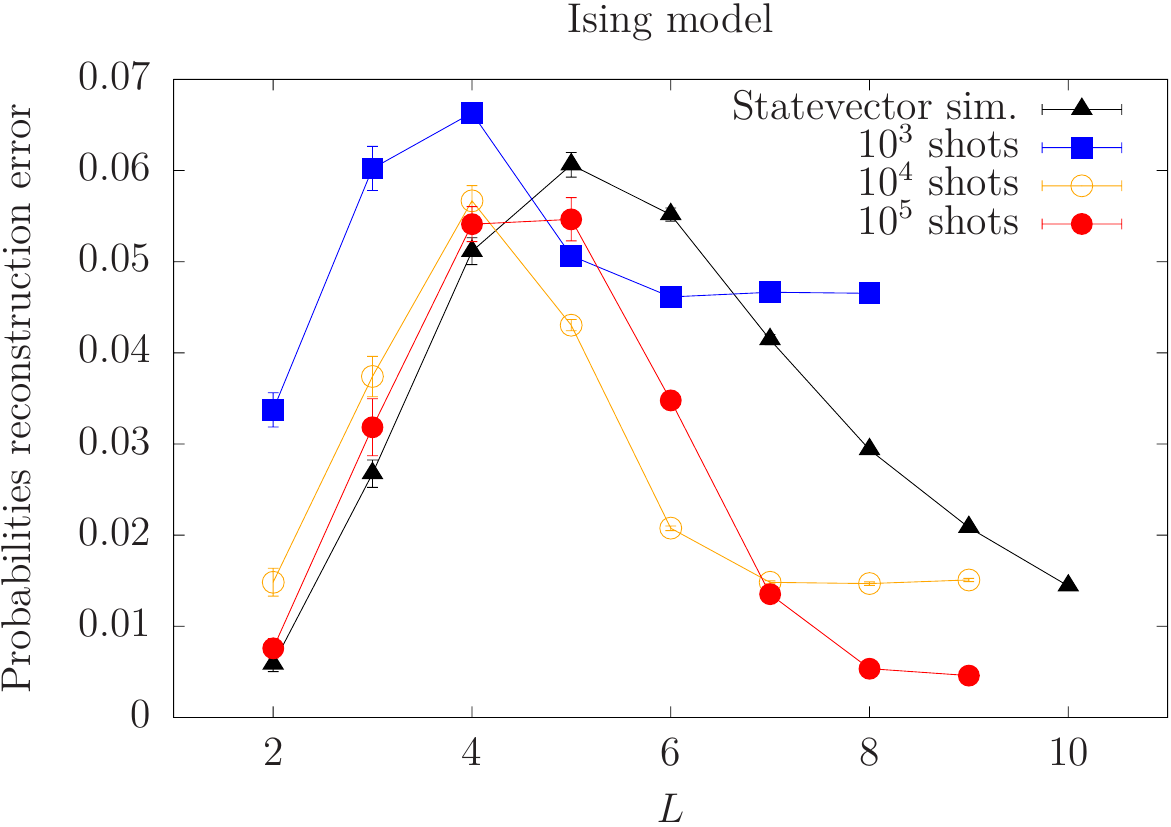}\caption{
Error $\epsilon_{{\bf b}1}$ in the $P_{{\bf b}1}$ probability reconstruction from Eq.~\ref{eq:err}, as a function of the system size $L$, for the Ising model. 
Black points correspond to the state vector simulations where we can exactly generate the probabilities $P_{\bf b}$ and $\bar P_{\bf b}$.
The error does not grow indefinitely with $L$, and it eventually  decreases.
The procedure is robust against discretization of probabilities, which are reconstructed through measurements, using $10^3, 10^4, 10^5$ shots (blue, orange and red datasets respectively).
The finite sampling prevents an exact solution of the system of the minimization problem in Eq.~\ref{eq:opti}, nevertheless the error saturates at a finite (small) value.
}
\label{fig:rec_size}
\end{figure}

\begin{figure}[hbt]
\includegraphics[width=1\columnwidth]{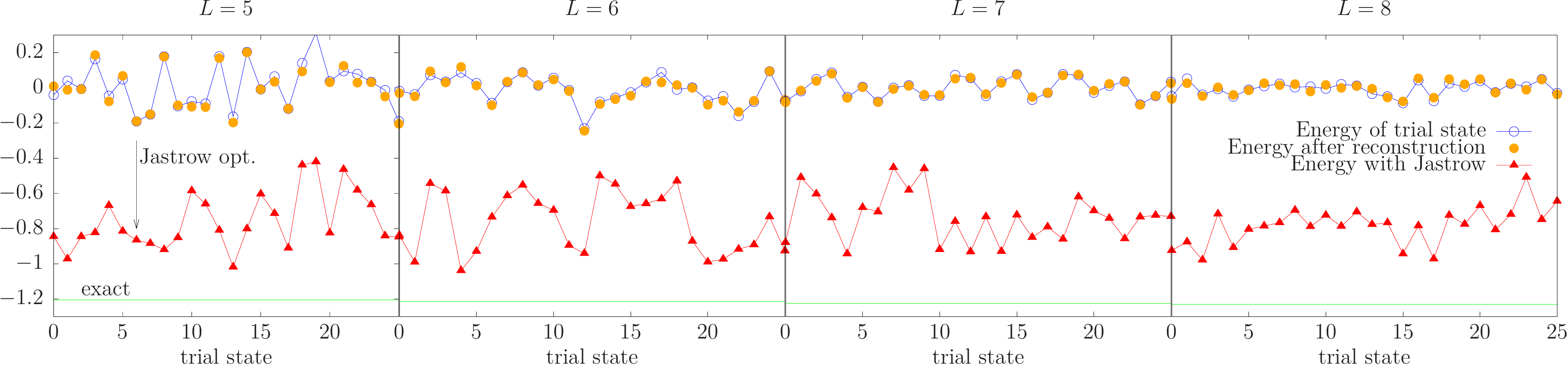}\caption{
We generate random real-valued states for each given system size $L$ (shown are 25 states with $L$ ranging from $L=5$ to $L=8$) and we compute the energy without the entangled copy circuit (the energy of the random state, blue), the energy obtained using the entangled copy circuit and reconstructing the  probabilities as in Eqs.~\ref{eq:rec},\ref{eq:modif},\ref{eq:opti} (orange), then the energy obtained using also the Jastrow parameters (optimizing only the Jastrow, at fixed $\bar P_{\bf b}$, red).
We observe that the energy reconstruction works quite well, but it does not need to be optimal since the measurements are reweighted anyway using the Jastrow optimization.
In green we plot also the exact energy of the model, showing that the method is variational. In this run it is quite unlikely to recover the exact energy since the Jastrow projector cannot chang e the signs of the components of the random state.
Nevertheless the accuracy improvement is evident, and does not degrade as the system size increase.
Data is uncorrelated, so lines here represent only a guide for the eye.
}
\label{fig:rec_ene}
\end{figure}

However we can check that, for the Ising (shown in fig.~\ref{fig:rec_size}) and Heisemberg model the error in the distribution defined as 
\begin{equation}
\label{eq:err}
  \epsilon_{\bf b}= \left( \sum_i | P_{\bf b}(i) -  P^0_{\bf b}(i) |^2 \right)^{1/2}
\end{equation}
is finite and small, and the resulting difference in the energies computed is negligible compared to the energy gain provided by Jastrow optimization (cfn.~\ref{fig:rec_ene}).

\clearpage
\section{Probabilities reweighting using the Jastrow operator}

Each measured probability $\bar P_{\bf b}$ in basis ${\bf b}$  is reweighted according to
\begin{equation}
     \bar P_{\bf b}(j ~2^L + i) \rightarrow \bar P_{\bf b}(j ~2^L + i) ~ w(j),
\end{equation}
where $w(j)$ is the Jastrow operator (which is calculated from  the ancilla register read-out), 
\begin{equation}
   \mathrm{exp}( \sum_{s,t=1 (s\ne t)}^N \lambda_{st} \sigma^z_s \sigma^z_t),
\label{eq:jastrow}
\end{equation}

and the correlator $\sigma^z_s \sigma^z_t$ is evaluated only on the ancilla bit-string $j$, where the qubits are measured in their Z basis.
The probability is then renormalized to 1, and is processed  to give the reduced probability $ P_{\bf b}$ as explained in the above section.

\clearpage
\section{Energy fluctuations due to measurements reweighting}

Poor performances are expected when the circuit state $\psi_c$ has negligible overlap with the exact one, which is the solution of the problem Hamiltonian that we aim to minimize.
While numerically this issue is manifested by a vanishing normalization of the JQC state, in the proposed implementation $A$ based on measurement reweighting, an increased statistical fluctuation of the energy values would be a fingerprint that most of the measurements are reweighted to zero, with only very few samples, collected from the circuit measurements, being enhanced accordingly by the Jastrow factor.

A qualitatively correct circuit ansatz is therefore of central importance for the method. We show that, despite its simplicity, the heuristic ansatz here considered maintain a good overlap with the exact state as the system size $L$ is increased.
In this way the number of components projected-out by the Jastrow operator stays relatively constant with increasing system size.
The observed dispersion of the energies as a function of the Jastrow parameters amplitude is therefore also approximately constant.

\begin{figure}[hbt]
\includegraphics[width=0.5\columnwidth]{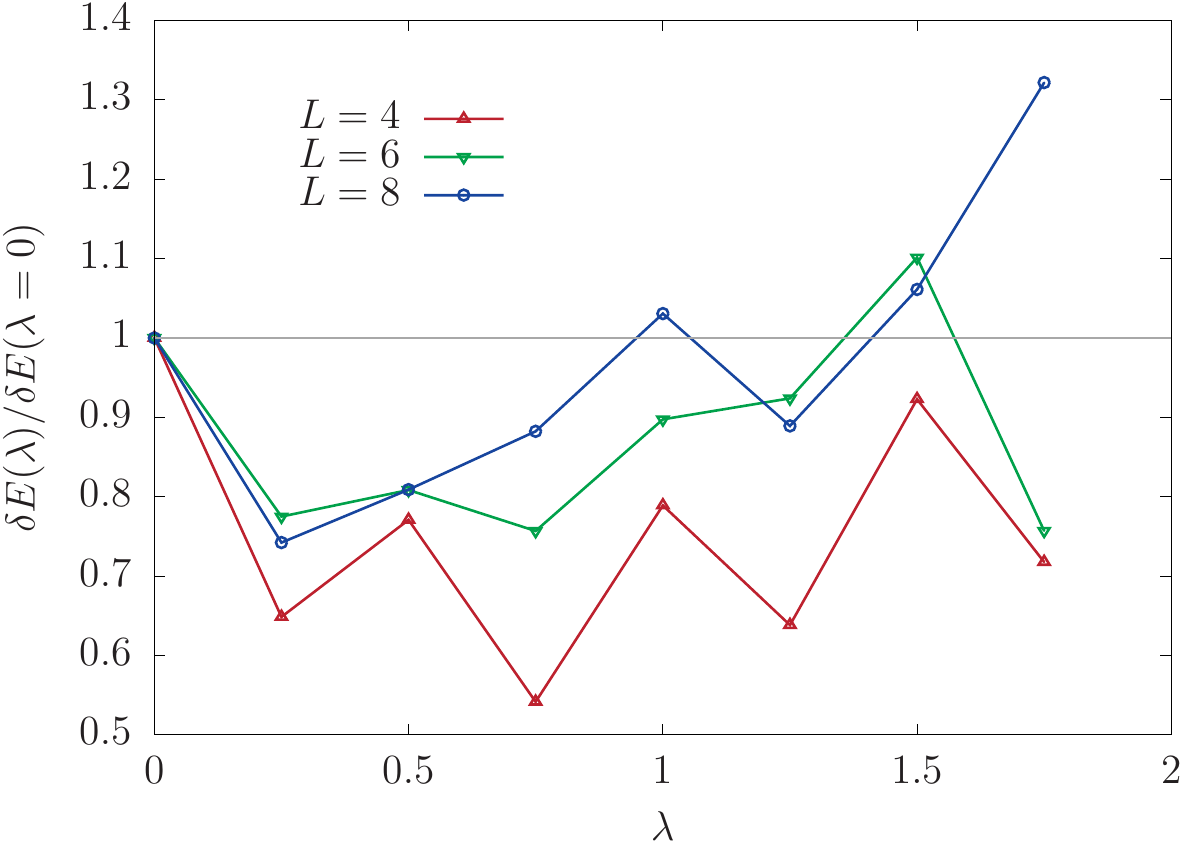}\caption{
Dispersion of the energy values computed by repeating the process $M_{rep}=12$ times. Each energy evaluation has been obtained using 320000 $L$-qubits measurements in the Z and X bases.
The model considered here is the Ising model.
The $\lambda$ value is  controls an element-wise reparametrization of the optimal Jastrow parameters, such that when $\lambda=0$ the JQC state reduces to the circuit state $\psi_c$ and no measurement reweighting is applied, whereas for  $\lambda=1$ the optimal Jastrow parameters are recovered and the JQC state has the minimum energy.
}
\label{fig:hubbb}
\end{figure}

\section{Implementation $B$}

As discussed in the main text, another viable option to implement effectively non-unitary operators is by computing the 
 the ratio of the expectation values of the operators $\mathcal{P}_J \mathcal{H} \mathcal{P}_J$ and $\mathcal{P}_J \mathcal{P}_J$ on $\psi_c$.
 This means measuring different Pauli operators. 
 Unfortunately, it can be verified that the exponential ansatz of the projector (cfn. Eq. 1 of the main text)
 \begin{equation}
     \mathcal{P}= e^J
     \label{eq:aaa}
 \end{equation} 
 implies a number of Pauli terms that grows exponentially with the system size $N$. This is true already when expressing the normalization operator $\mathcal{P}_J \mathcal{P}_J$ as sum of Pauli strings.
 
 A possible workaround is to truncate the exponential ansatz using
 \begin{equation}
     \mathcal{P}= (1 + J)^s~.
     \label{eq:bbb}
 \end{equation}
 
\begin{figure}[hbt]
\includegraphics[width=0.5\columnwidth]{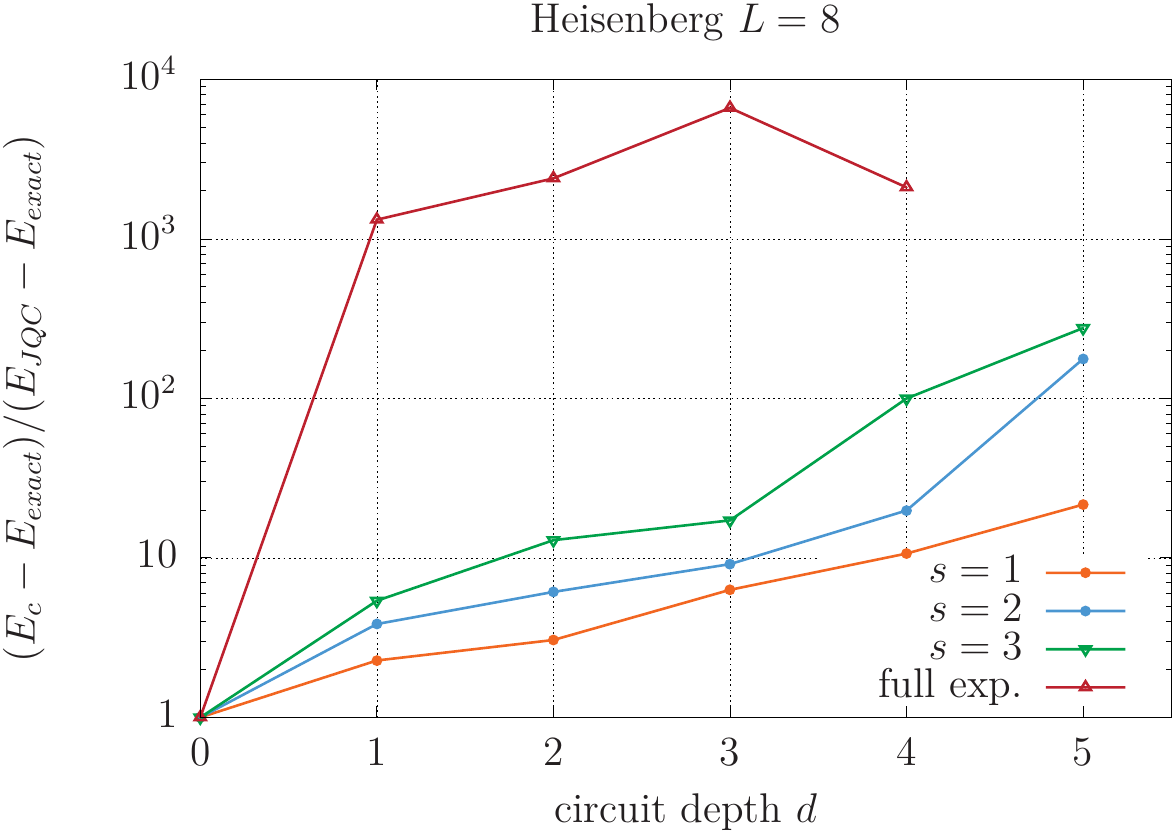}\caption{Computational gain provided by the Jastrow operator as a function of the circuit depth. This is an $R_y$-CNOT heuristic circuit (cfn. Fig.~\ref{fig:rycnot})
}
\label{fig:2impl}
\end{figure}

 With this choice the number of Pauli strings is polynomially increasing with $N$.
 We tested analytically how the truncation affects the quality of the ansatz. Results are shown in Fig.~\ref{fig:2impl} in the case of the $L=8$ Heisemberg model.
 We see that for low circuit depths $d$ the computational gain is much more limited compared to the exponential ansatz considered in the main text and previous section of this Supplementary Materials.
 Given the relatively small computational gain, and the necessity of perfoming more measurements, we focused on the \emph{implementation $A$} in the main text.

\end{document}